\documentclass[aps,prd,twocolumn,amsmath,floatfix,longbibliography,superscriptaddress,preprintnumbers,
]{revtex4-1}
\usepackage{lineno,isotope,mathrsfs}
\modulolinenumbers[5]
\usepackage{amsmath,bbold,amssymb,epsfig,bm,mathrsfs,feynmp}
\usepackage{color,slashed,nicefrac,exscale,multirow,soul}
\usepackage{times,txfonts,xcolor,graphicx,gensymb,tikz}
\usepackage[normalem]{ulem}
\usepackage[breaklinks,colorlinks,citecolor=blue]{hyperref}
\definecolor{red}{rgb}{0.8,0,0}
\definecolor{violet}{rgb}{0.4,0,0.4}
\definecolor{green}{rgb}{0,0.5,0.0}
\definecolor{navy}{rgb}{0.0,0.0,0.6}
\definecolor{orange}{rgb}{0.8,0.2,0.0}

\usepackage[normalem]{ulem}  

\newcommand{\bea}{\begin{eqnarray}}
\newcommand{\eea}{\end{eqnarray}}

\newcommand{\vecp}{{\bm p}}

\newcommand{\vecv}{{\bm v}}
\newcommand{\vecE}{{\bm E}}

\newcommand{\vecj}{{\bm j}}
\newcommand{\vecB}{{\bm B}}

\newcommand{\ie}{{\it i.e.}}

\input{acronym.input}

\begin{document}
\title
{Electrical conductivity of a warm neutron star crust in magnetic fields: Neutron-drip regime
}

\author{Arus Harutyunyan}
\email{arus@bao.sci.am}
\affiliation{Byurakan Astrophysical Observatory,
  Byurakan 0213, Armenia}
\affiliation{Department of Physics,
                  Yerevan State University,
                  Yerevan 0025, Armenia}
\author{Armen Sedrakian}
\email{sedrakian@fias.uni-frankfurt.de}
\affiliation{Frankfurt Institute for Advanced Studies,
D-60438 Frankfurt am Main, Germany}
\affiliation{Institute of Theoretical Physics,
University of Wroclaw, 50-204 Wroclaw, Poland}                                                                              
\author{Narine T. Gevorgyan} \email{gevorgyan.narine@gmail.com }
\affiliation{Byurakan Astrophysical Observatory, Byurakan 0213,
  Armenia} \affiliation{A. I. Alikhanyan National Science Laboratory
  (Yerevan Physics Institute), Yerevan 0025, Armenia}
\author{Mekhak V. Hayrapetyan}
\email{mhayrapetyan@ysu.am}
\affiliation{Department of Physics,
                  Yerevan State University,
                  Yerevan 0025, Armenia}
\begin{abstract}
  We compute the anisotropic electrical conductivity tensor of the
  inner crust of a compact star at non-zero temperature by extending a
  previous work on the conductivity of the outer crust. The physical
  scenarios, where such crust is formed, involve proto-neutron stars
  born in supernova explosions, binary neutron star mergers and
  accreting neutron stars. The temperature-density range studied
  covers the transition from a semi-degenerate to a highly degenerate
  electron gas and assumes that the nuclei form a liquid, \ie, the
  temperature is above the melting temperature of the lattice of
  nuclei. The electronic transition probabilities include (a) the
  screening of electron-ion interaction in the hard-thermal-loop
  approximation for the QED plasma, (b) the correlations of the ionic
  component in a one-component plasma, and (c) finite nuclear size
  effects.  The conductivity tensor is obtained from the Boltzmann
  kinetic equation in relaxation time approximation accounting for the
  anisotropy introduced by a magnetic field.  The sensitivity of the
  results towards the matter composition of the inner crust is
  explored by using several compositions of the inner crust which were
  obtained using different nuclear interactions and methods of solving
  the many-body problem.  The standard deviations of relaxation time
  and components of the conductivity tensor from the average are below
  $\le 25\%$ except close to crust-core transition, where
  non-spherical nuclear structures are expected. Our results can be
  used in dissipative magnetohydrodynamics simulations of warm compact
  stars.

\end{abstract}
\preprint{INT-PUB-23-027}

\maketitle
%
\section{Introduction}
\label{sec:Intro}

The knowledge of transport properties of hot baryonic matter is
important for large-scale magneto-hydrodynamics description of
astrophysical phenomena associated with compact stars. One such
setting offer the binary neutron star (BNS) mergers, such as the
GW170817 event~\cite{LIGO-Virgo:2019}: matter is expected to be heated
both in the post-merger and pre-merger phases. In the post-merger
phase, the matter is heated to temperatures up to 100-150~MeV by
deposition of kinetic and gravitational energy in the matter; in the
pre-merger phase, the matter may be heated via the dissipation of the
energy of the tidally induced oscillations. Another longer time-scale
setting is offered by the accreting neutron stars as their crusts are
heated through the infalling matter and the onset of nuclear reactions
in various crustal layers. The warm matter regime is of interest also
in the context of transient proto-neutron stars born in supernova
explosions.  {In this work we focus on stellar matter at subnuclear
  densities at moderate temperatures in the range
  $T_m\le T\le T_{\rm tr}$, where $T_m \simeq 1$~MeV is the melting
  temperature of the crustal lattice corresponding to matter featuring
  heavy nuclei, dripped neutrons, and relativistic electron gas in
  liquid matter regime and $T_{\rm tr}\simeq 5$~MeV is the trapping
  temperature of neutrinos.}

Transport in compact star plasma has been extensively studied in the
cold and dense limit where it is dominated by the degenerate fermionic
quantum liquids over a long period; for general reviews see
Refs.~\cite{Schmitt2018, Potekhin2018}. More recently, the electrical
conductivity of the warm outer crust was computed in
Ref.~\cite{Harutyunyan2016} in the context of BNS mergers; it also
provides a detailed review of the previous work on the conductivity of
crustal matter in the cold regime, which we do not repeat here. The
conductivity tensor of Ref.~\cite{Harutyunyan2016} was then used to
assess its importance in the dynamics of BNS
mergers~\cite{Harutyunyan2018}, showing the conditions for the
breakdown of the ideal magnetohydrodynamics (MHD) limit and the
importance of the Hall conductivity.

This paper extends a previous calculation of the conductivity of
heated crustal matter in non-quantizing magnetic
fields~\cite{Harutyunyan2016} to the inner crust phase where along
with the crustal lattice, there is an unbound neutron component. Our
focus is on the case with spherical nuclei for which several
compositions~\cite{Negele1973,Mondal2020,Pearson:2018,Raduta2019} will
be used.  {Our collection includes two models with zero-range Skyrme
  interaction~\cite{Pearson:2018,Raduta2019}, which use modern and
  accurate parametrizations of this force, see
  also~Ref.~\cite{Chamel2011}.  The two models of
  Ref.~\cite{Mondal2020} use modern parametrizations of the
  finite-range Gogny interaction. We also included the widely used
  model of Ref.~\cite{Negele1973} which is based on a
  density-dependent effective Hamiltonian tuned on finite nuclei; see
  Sec.~\ref{sec:regimes} for a further discussion of these models. }
The conductivity of pasta phases where the shapes and topologies of
nuclei deviate strongly from the spherical within the layer between
the stellar core and the phase with spherical nuclei was studied in
Ref.~\cite{Pelicer2023}. { The low-temperature regime of electrical
  and thermal conductivity, where the nuclei form a lattice and
  interactions are mediated by lattice phonons was studied in
  Ref.~\cite{Itoh:1984a}}.  We will restrict our discussion to
magnetic fields below $B_c \simeq 10^{14}$~G. Above these values, the
Landau quantization of electron trajectories must be taken into
account, see Ref.~\cite{Potekhin1999}.

The paper is organized as follows. In Sec.~\ref{sec:regimes} we review
the compositions of the neutron star's inner crust, which are used in
our computations of the electrical conductivity
tensor. Section~\ref{sec:Conductivity} collects the relevant
ingredients of the formalism and key results, which have been
presented in detail in Ref.~\cite{Harutyunyan2016}. Our numerical
results are discussed in Sec.~\ref{sec:Results}. The final
Sec.~\ref{sec:Conclusions} contains a summary of our results.

We use the natural (Gaussian) units with $\hbar= c = k_B = k_e = 1$,
$e=\sqrt{\alpha}$, $\alpha=1/137$, and the metric signature
$(1,-1,-1,-1)$.

\section{Equation of state, composition and physical conditions in inner crust}
\label{sec:regimes}

Above the neutron drip density $\rho_{\rm drip} = 4.3\times 10^{11}$ g
cm$^{-3}$ a phase transition takes place in neutron star crusts: the
low-density phase consisting of fully ionized nuclei, which are
characterized by the nucleon number $A$ and proton number $Z$, and
relativistic electrons is replaced by a phase which in addition to the
components of the low-density phase has also unbound neutrons.  Then,
the total baryon density $n_B$ is given by
\bea\label{eq:n_B}
n_B = An_i + n_n',
\eea
where $n_i$ is the number density of the ions (nuclei) {and
  $n_n'= (1-V_Nn_i)n_n$, where $n_n$ is the number density of unbound
  neutrons, $V_N$ is the volume of the nucleus, and the term $V_Nn_i$
  is the excluded volume correction~\cite{Baym1971}.}  The
ion-electron sub-system, viewed as Coulomb plasma, is characterized by
the parameters
\bea\label{eq:Gamma}
\Gamma= \frac{T_C}{T}, \qquad T_C = \frac{e^2 Z^2}{a_i},
\eea
where $e$ is the elementary charge, $T$ is the temperature,
$a_i=(4\pi n_i/3)^{-1/3}$ is the radius of the spherical volume per ion, \ie, that of the Wigner-Seitz cell. For $\Gamma\ll 1$ ($T\gg T_{\rm C}$) ions are weakly coupled and
because of their large mass they form a classical Boltzmann gas.  When
$\Gamma\ge 1$ ions are strongly coupled and form a solid phase with
nuclei arranged in a regular lattice for $\Gamma>\Gamma_m \simeq 160$.
In the opposite case $\Gamma<\Gamma_m$ the liquid phase is
energetically preferred. The temperature of melting of the crustal
lattice is given by $T_m=T_C/\Gamma_m$. The lattice plasma
temperature is defined as
\bea 
T_p = \biggl(\frac{4\pi  Z^2e^2n_i}{M }\biggl)^{1/2} ,
\eea
where $M $ is the ion mass. (Note that in units  where $\hbar=1$,
the plasma frequency and plasma temperature coincide). 
The quantum regime for ion lattice (under which the quantization of
oscillations of the lattice is required) occurs for
$T\le T_p$.  

\begin{figure}[bht] 
\begin{center}
\includegraphics[width=\linewidth,keepaspectratio]{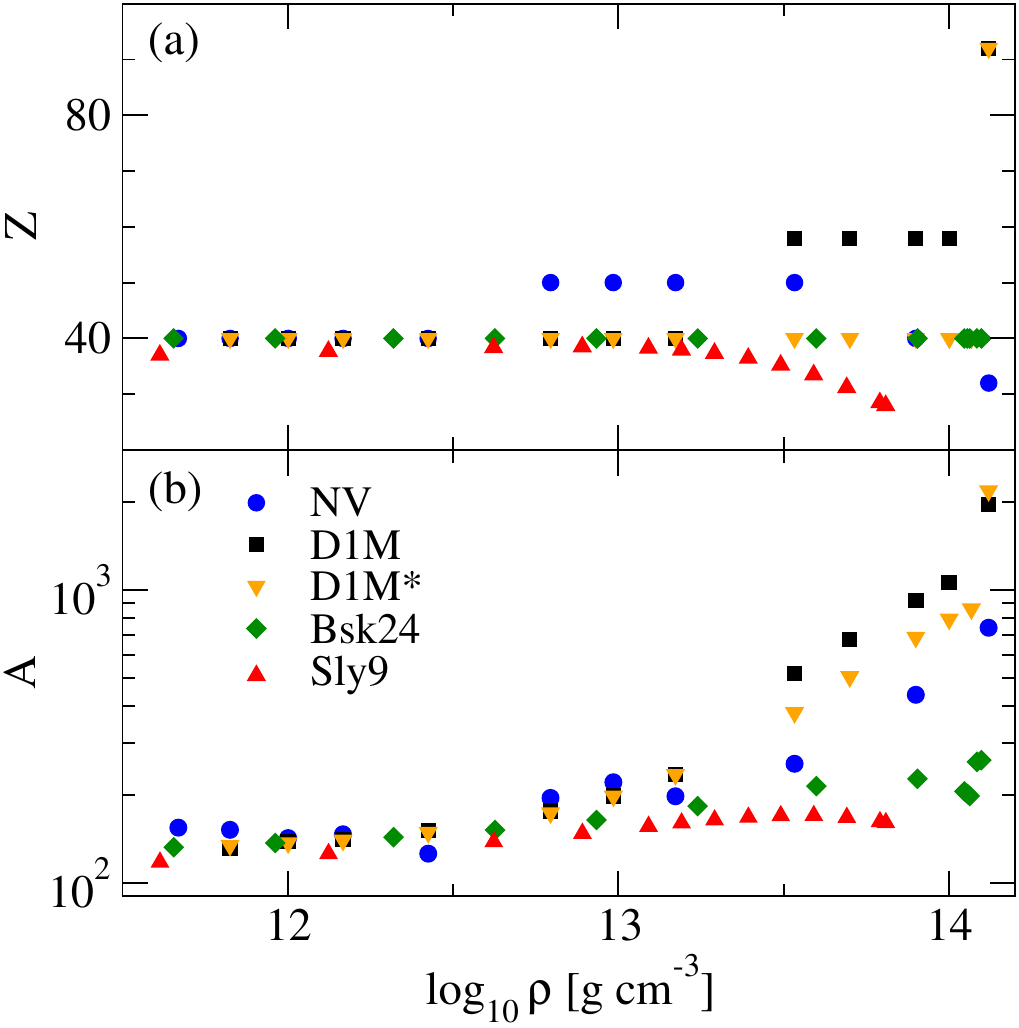}
\caption{The proton number $Z$ (top panel) and the nucleon number
  $A$ (bottom panel) of the nuclei as functions of the mass density for
  five different compositions of the stellar matter labeled as
  NV~\cite{Negele1973}, D1M and D1M$^*$~\cite{Mondal2020},
  Bsk24~\cite{Pearson:2018}, and Sly9~\cite{Raduta2019}.  }
\label{fig:compositions} 
\end{center}
\end{figure}
\begin{figure}[tbh] 
\begin{center}
\includegraphics[width=\linewidth,keepaspectratio]{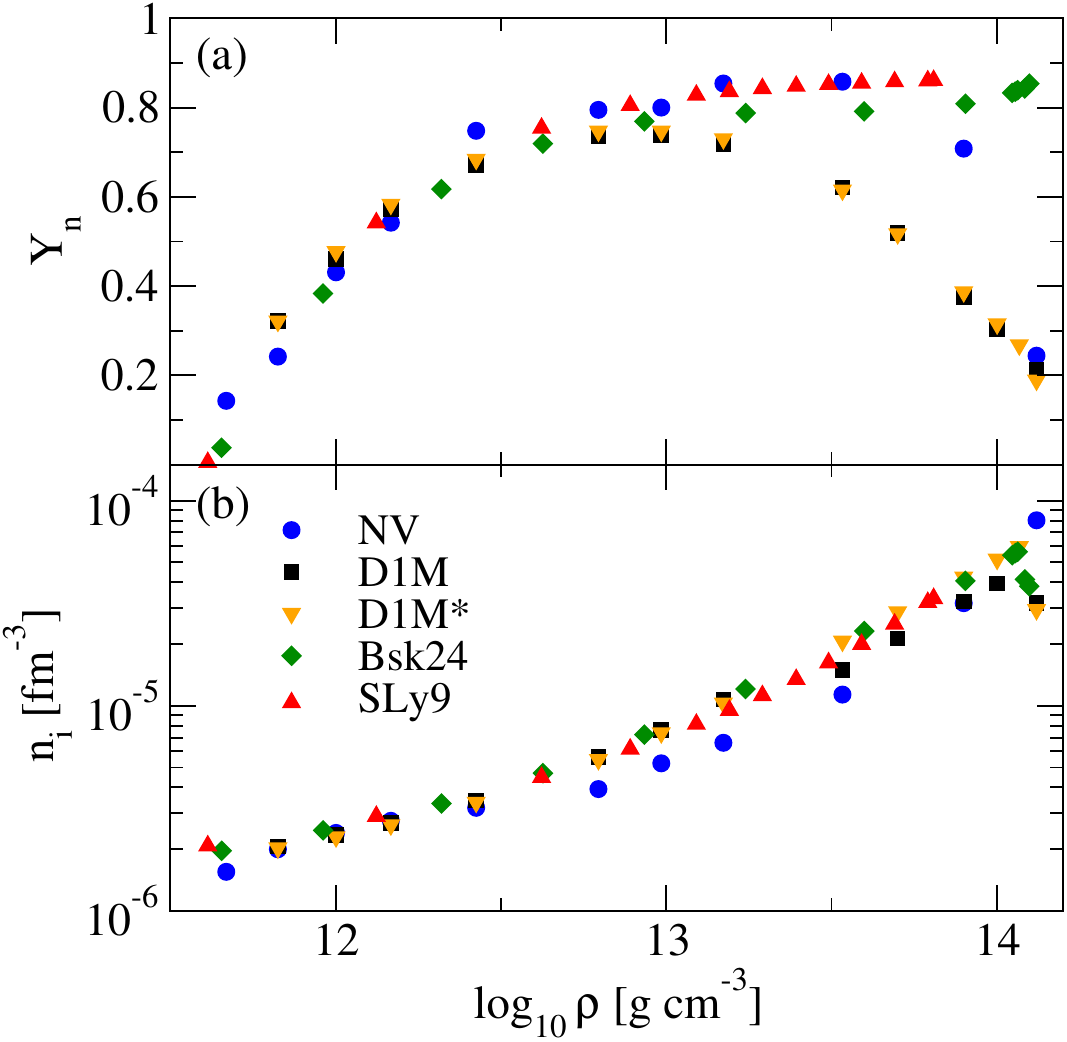}
\caption{(a) The fraction of free neutrons $Y_n=n'_n/n_B$ and
  (b) the number density of ions (b) as functions of the mass density for five compositions of stellar matter. 
}
\label{fig:neutrons} 
\end{center}
\end{figure}

For numerical computations, we will adopt five different
density-dependent compositions of stellar matter for the inner crust
of a neutron star, which we label as NV~\cite{Negele1973}, D1M and
D1M$^*$~\cite{Mondal2020}, Bsk24~\cite{Pearson:2018}, and
Sly9~\cite{Raduta2019}. These compositions were
computed at $T=0$. 
{
  Reference~\cite{Negele1973} starts with the density-dependent
  Hartree-Fock theory and performs density matrix expansion to write
  down a Hamiltonian, which depends on the densities of protons and
  neutrons and their gradients. The model of the inner crust is then
  obtained by minimizing the energy per nucleon of spherically
  symmetric configurations of nucleons in a Wigner-Seitz unit cell
  with a uniform background gas of degenerate electrons. The pairing
  correlations and shell effects were neglected. Despite its
  limitations, the model of Ref.~\cite{Negele1973} has been used in
  astrophysical applications extensively, in particular, in the
  studies of transport in the solid regime at low
  temperatures~\cite{Itoh:1984a}.
  The two models of Ref.~\cite{Mondal2020} were obtained using
  finite-range Gogny interactions to obtain the nuclear mean field and
  pairing correlations self-consistently. The model of the inner crust
  was obtained using the semiclassical variational Wigner-Kirkwood
  method, which incorporated both shell and pairing corrections, which
  were, respectively, computed using the Strutinsky integral and the
  Bardeen-Cooper-Schrieffer (BCS) mean-field approximation.
  The model of Ref.~\cite{Pearson:2018}, see also
  Ref.~\cite{Chamel2011}, uses zero-range Bsk24 Skyrme functional from
  the family of Brussels–Montreal nuclear density functionals and
  computes the properties of the inner crust using the
  temperature-dependent extended Thomas-Fermi method, supplemented
  with Strutinsky integral method to account for shell effects and BCS
  pairing through inclusion of BCS pairing energy in the density
  functional.
  Reference~\cite{Raduta2019} uses zero-range Skyrme functional for unbound
  particles and beyond drip-line nuclei and realistic nuclear mass and
  level density tables for known nuclei in a model of a statistical
  distribution of Wigner–Seitz cells. It includes phenomenological
  pairing correction term that scales with mass number and the shell
  effects are accounted for automatically in the experimental data
  used.  }

We will assume below that the composition does not
depend strongly on the temperature in the range of temperatures
studied here so that the background composition of the inner crust in each case will be fixed at the one derived for $T=0$.  The physical
conditions change significantly with the increasing temperature at
about $T_{\rm tr}\simeq 5$~MeV where the neutrinos become trapped and
$\beta$-equilibrium conditions are changed. As the temperature is
further increased the appearance of lighter clusters -- deuterons,
tritons, helions, and alpha particles become possible for temperatures
$T\ge
10$~MeV~\cite{Typel2010PhRvC,Hempel2011PhRvC,Hempel2012ApJ,Wu2017LTP,Fischer2020PhRvC,Sedrakian2020EPJA}.
Their contribution to conductivity is left for future study.

Figure~\ref{fig:compositions} shows the proton number $Z$ and the
nucleon number $A$ of the nuclei as functions of the net mass density
for the chosen compositions of the stellar matter. Up to densities
$\log_{10} \rho\, [\textrm{g\, cm}^{-3}] \le 13$ all predict $Z=40$
semi-magic proton values.  Above these densities, Bsk24 and D1M$^*$
predict the same $Z=40$ value, NV composition predicts the magic value
$Z=50$ (except the last point) whereas D1M and Sly9 predict higher and
lower $Z$ values, respectively.  Therefore, the matrix elements for
the electron scattering off the individual nuclei are nearly the same
for these compositions at low densities but deviate at higher
densities.  However, since the transport depends on the average number
of nuclei per unit volume, such factors as the free neutron density
and the mass number of a nucleus are important, see
Eq.~\eqref{eq:n_B}, which non-trivially modify the predictions based
on the value of $Z$.

Figure~\ref{fig:neutrons} shows the fraction of free neutrons
$Y_n=n'_n/n_B$ defined as the ratio of the number of free neutrons to
that of all nucleons in the Wigner-Seitz cell and the ion number
density as functions of the mass density for all five compositions
studied. Notable deviations are seen in free neutron fractions in the
high-density range $13\le \log_{10}\rho$ [g~cm$^{-3}]\le 14$, where
the models D1M and D1M$^*$ predict decreasing free neutron fractions
because of the fast increase of the nuclear mass number $A$ at these
densities. In the models Bsk24 and Sly9 $Y_n$ remains almost constant
($Y_n\simeq 0.8$) at $\log_{10}\rho$ [g~cm$^{-3}]\ge 13$, whereas the
model NV shows somewhat intermediate behavior between these two model
types at $\log_{10}\rho$ [g~cm$^{-3}]\ge 13.5$ where the neutron
fraction drops again.

\begin{figure}[t] 
\begin{center}
\includegraphics[width=\linewidth,keepaspectratio]{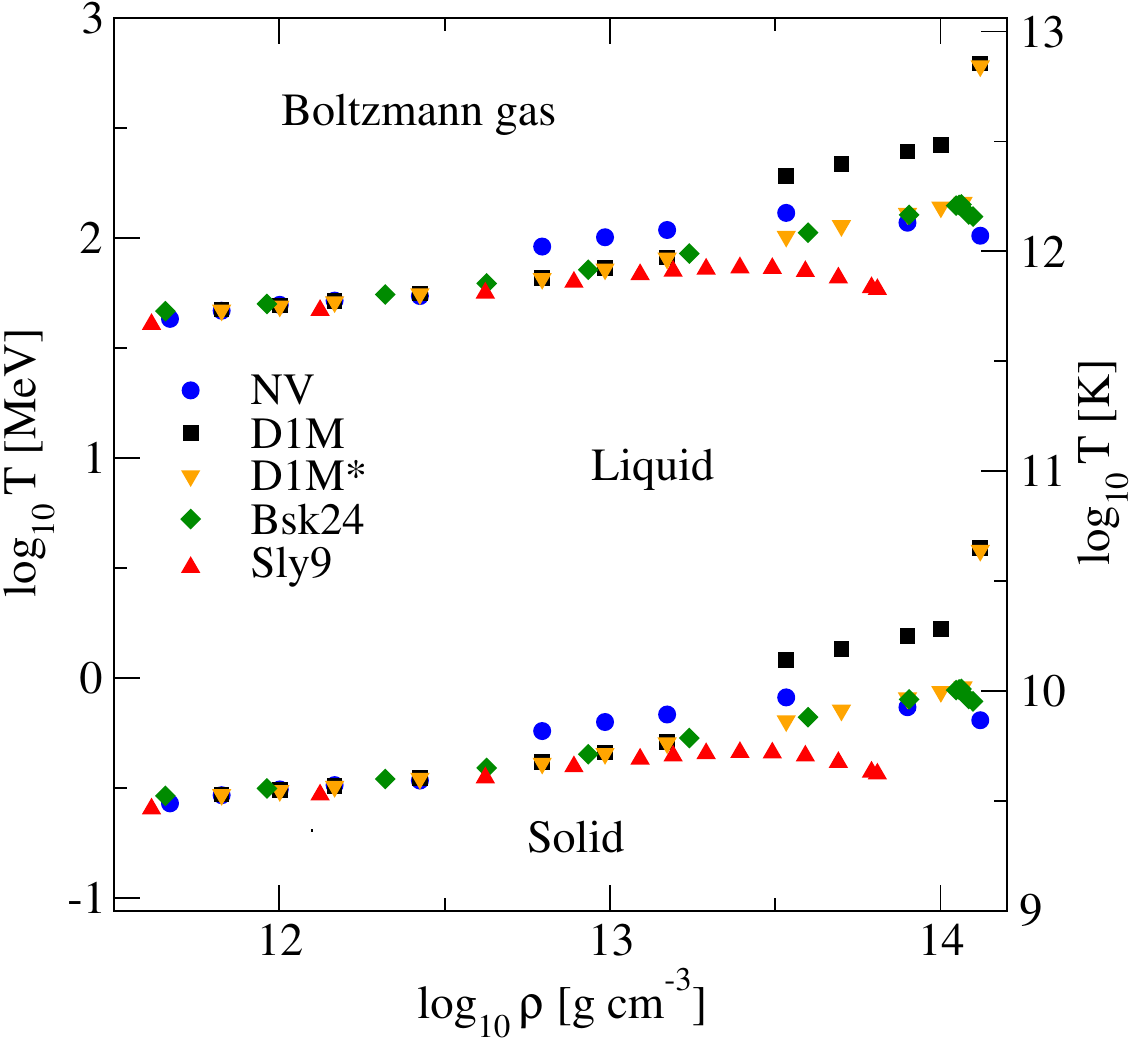}
\caption{ The phase diagram of dense plasma in the inner crust of 
the neutron star in the temperature-density plane for five different
  compositions.  The lower
  curves show the melting temperature $T_m$ below which the ionic
  component solidifies. 
  Upper curves show
  $T_{\rm C}$ above which the ionic component forms a Boltzmann
  gas. The present study covers the liquid portion of the phase
  diagram. }
\label{fig:PhaseDiagram} 
\end{center}
\end{figure}

Figure \ref{fig:PhaseDiagram} shows the phase diagram of the matter in
the inner crust of neutron stars in the temperature-density plane
for five compositions. In the top part of the diagram where $T> T_C$ the ionic
component forms a weakly interacting Boltzmann gas, as the thermal
energy exceeds the Coulomb interaction energy. In the bottom part of
the diagram where $T< T_{m}$ the ionic component solidifies, \ie, the
scattering of electrons is (predominantly) on the phonons of the lattice. The plasma
temperature $T_p$ is lower than the melting temperature for the five
models in Fig.~\ref{fig:PhaseDiagram} and is not shown.  Our results
apply in the regime where $T>T_{m}$. Note the weak density and model
dependence of the curves $T_{m}(\rho)$ and $T_{C}(\rho)$
(except the region $\log_{10}\rho$ [g~cm$^{-3}]\ge 13.5$). 
\begin{figure}[!] 
\begin{center}
\includegraphics[width=\linewidth,keepaspectratio]{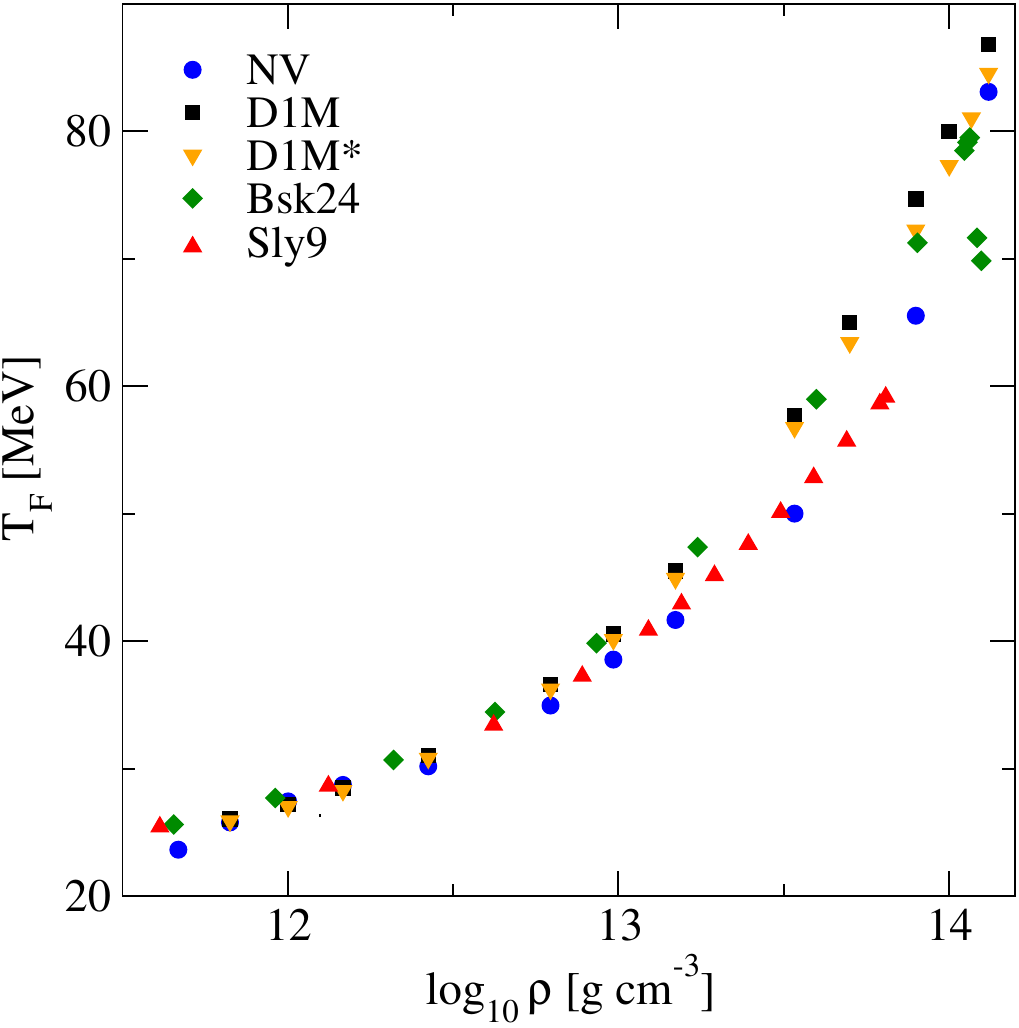}
\caption{ The Fermi temperature $T_F$ of the electronic component of
  the stellar matter in the inner crust of a neutron star for five
  different compositions shown in Fig.~\ref{fig:compositions}. The
  electron gas is becoming gradually non-degenerate above and
  degenerate below this temperature.}
\label{fig:TempFermi}
\end{center}
\end{figure}

The transport in the liquid phase of a neutron star's inner crust is
controlled by the electrons, which, to a good approximation, can be
treated as a free Fermi gas except for the collisions with ions that
lead to dissipation contributing to conductivity. (Electron-electron
interaction can affect the conductivity indirectly by modifying the
density of state of electrons, but the correction is a higher-order effect
in the fine structure constant). The electron density is obtained from
the charge conservation $n_e=Zn_i$ and allows us to define their
low-temperature characteristics, such as Fermi energy
$\varepsilon_F= (p_F^2+m^2)^{1/2}$ and temperature
$\varepsilon_F-m\equiv T_F$, where $m$ is the electron mass and
$p_F = (3\pi^2n_e)^{1/3}$ is the Fermi momentum. Note that at
  densities considered in this paper electrons are ultrarelativistic
 therefore we have, in practice, $T_F=\varepsilon_F=p_F$.

 The Fermi temperature for five compositions is shown in
 Fig.~\ref{fig:TempFermi}.  It is seen that electrons are degenerate
 up to a quite high temperature of several tens of MeV. This is in
 contrast to the outer crust region~\cite{Harutyunyan2016}, where low
 densities required treatment of the transition from degenerate to the
 non-degenerate regime. Nevertheless, as our set-up works across from
 non-degenerate to strongly degenerate limits, no additional
 limitations are imposed.  Overall, we
 see that electrons are degenerate or semi-degenerate for the
 temperatures relevant to BNS mergers and core-collapse supernovas at
 densities relevant to compact star inner crust.  We recall that our
 discussion is limited to temperatures below several MeV, as the
 compositions adopted will be modified due to finite-temperature
 effects at larger temperatures. In addition, to modifications of
 thermodynamics, additional species such as 
 alpha particles and other light clusters will appear in the
matter~\cite{Typel2010PhRvC,Hempel2011PhRvC,Hempel2012ApJ,Wu2017LTP,Fischer2020PhRvC,Sedrakian2020EPJA}  and will contribute to the conductivity. 
As a consequence, we will focus on the degenerate electron regime while providing some
 extrapolation to higher temperatures which are suggestive of the
 behavior of various quantities.

\section{Conductivity in magnetic field from Boltzmann equation}
\label{sec:Conductivity}

In this section, we provide the key ingredients of the formalism
presented extensively in Ref.~\cite{Harutyunyan2016} which is based on
the (quasi)particle transport of electrons in strong magnetic
fields. The kinetics of electrons is described by the Boltzmann equation for the electron distribution function
\bea\label{eq:boltzmann1}
\frac{\partial f}{\partial t}+
\bm v\frac{\partial f}{\partial\bm r}-
e(\vecE+[\vecv\times \vecB])\frac{\partial f}
{\partial\vecp}=I[f],
\eea
where $\vecE$ and $\vecB$ are the electric and magnetic fields,
$\vecv$ is the electron velocity, $e$ is the unit charge, and $I[f]$
is the collision integral, which for electron-ion collisions has
the form
\bea\label{eq:collision1}
I&=&-(2\pi)^4\sum\limits_{234}|{\cal M}_{12\to 34}|^2\delta^{(4)}(p+p_2-p_3-p_4)\nonumber\\
&\times&
[f(1-f_3)g_2-f_3(1-f)g_4],
\eea
where $f=f(p)$ and $f_3=f(p_3)$ are the distribution functions of the
incoming and outgoing electron, $g_{2,4}=g(p_{2,4})$ are the
distribution functions of the ion before and after a collision, and we
introduced the short-hand notation:
$\sum\limits_{i}=\int d\vecp_i/(2\pi)^3$.  As discussed above, ions
form a classical liquid in equilibrium with the
Maxwell-Boltzmann distribution, \ie,
\bea\label{eq:maxwell}
g(p)=n_i\,\bigg(\frac{2\pi}{MT}\bigg)^{3/2}
\exp\left({-\frac{p^2}{2MT}}\right).
\eea
The range of validity of the kinetic equation \eqref{eq:boltzmann1} 
follows from the common considerations of the kinetic 
theory~\cite{Lifshitz_Pitaevskii1981}.
Specifically, it is  assumed that the collisions are instantaneous, 
i.e., the distribution function is tracked only in between collisions.  
Similarly, we assumed that the range of scattering (effective size of screened Coulomb field $d$)  is much smaller than mean free path $\bar l$ between collisions. If the perturbations of the system  
are characterized by characteristic length $L$ and frequency $\Omega$ (for
example $L$ could be the scale set by gradients of electromagnetic potentials 
and $\Omega$ the frequency of oscillations) the general validity condition can be written as
 \bea 
 d\ll \bar l \ll L, \qquad \bar t\ll \Omega^{-1} ,\nonumber
\eea
where $\bar t$ is the mean-free-flight time associated with $\bar l$, which is  
of the order of the relaxation time of distribution function, see below.
Next, consider small perturbations $\delta f$ around the equilibrium Fermi-Dirac distribution function of electrons to linearize the Boltzmann equation: $f= f^0+\delta f$, $\delta f\ll f_0$, where the
equilibrium distribution is given by
\bea\label{eq:fermi}
f^0({\varepsilon})=\frac{1}{e^{
(\varepsilon-\mu)/T}+1},
\eea
with the spectrum of non-interacting electrons given by
$\varepsilon=\sqrt{p^2+m^2}$, and
$\mu$ is the electron chemical potential.  Since we are interested in
the electrical conductivity we keep only the last term on the
left-hand side of Eq.~\eqref{eq:boltzmann1}.  Substituting
$f= f^0+\delta f$ in Eq.~\eqref{eq:boltzmann1} and decomposing
$\delta f$ in terms of independent tensor components containing the
electric and magnetic fields we obtain~\cite{Harutyunyan2016}
\bea\label{eq:phi}
\delta f=\frac{e\tau}{1+(\omega_c\tau)^2}
 \frac{\partial f^0}{\partial\varepsilon}
v_i\left[\delta_{ij}-\omega_c\tau\epsilon_{ijk}
b_k+(\omega_c\tau)^2b_ib_j\right]E_j,\quad
\eea
where $\bm b \equiv \bm B/B$, $\omega_c=eB /\varepsilon$ is the cyclotron frequency for electrons, and the Latin indices label the components of Cartesian coordinates. 
Here we work in the relaxation-time approximation
with the relaxation time defined by
\bea\label{eq:t_relax}
\tau^{-1}(\varepsilon)&=&(2\pi)^{-5}\!\!
\int\!  d\bm q \!\int\! d\bm p_2\,
|{\cal M}_{12\to 34}|^2 \frac{\bm q\cdot \bm p}{p^2}
\nonumber\\ &\times &
\delta(\varepsilon+\varepsilon_2-\varepsilon_3-\varepsilon_4)
 g_2\frac{1-f^0_3}{1-f^0}.
\eea

The electrical conductivity is then obtained by computing the electrical current 
\bea\label{eq:current}
j_i=-2\!\!\int\!\frac{d\bm p}{(2\pi)^3}\,
ev_i\delta f =\sigma_{ij}E_j.
\eea 
Substituting Eq.~\eqref{eq:phi} in Eq.~\eqref{eq:current} and we find 
\bea\label{eq:sigma1}
\sigma_{ij}=\delta_{ij}\sigma_0-\epsilon_{ijm}b_m
\sigma_1 +b_ib_j\sigma_2,
\eea
where 
\bea\label{eq:sigma2}
\sigma_n=\frac{e^2}{3\pi^2T}\!\int_m^\infty\! d\varepsilon\,
\frac{p^3}{\varepsilon}\frac{\tau(\omega_c\tau)^n}
{1+(\omega_c\tau)^2}f^0(1-f^0),\quad n=0,1,2.
\nonumber\\
\eea

The matrix element includes several corrections to the bare Coulomb
interaction between an electron and an ion. The screening of the
interaction is taken into account via the hard-thermal-loop
polarization tensor of QED, see Sec.~IV D of
Ref.~\cite{Harutyunyan2016}. The ion-ion correlations are taken by
using fits to the structure factor $S(q)$ of one-component plasma for
various values of plasma parameter $\Gamma$ obtained from Monte Carlo
computations, see Fig.~4 of Ref.~\cite{Harutyunyan2016}. Finally, the
finite nuclear size of the nuclei is taken into account via a
nuclear form factor $F(q)$, which represents Fourier transform of 
the screened charge distribution of a spherically 
charged nucleus, see Eq.~(22) of Ref.~\cite{Itoh:1984a} or Eq.~(31) of
Ref.~\cite{Harutyunyan2016}. 
{The nuclear form factor $F(q)$ is evaluated with nuclear radii $R_N$ taken directly from underlying compositions, avoiding the relation $R_N=a A^{1/3}$  which is inaccurate for the inner crust compositions if $a=\textrm{const}$ is assumed.}

The final expression for the relaxation time reads
\begin{eqnarray}
\label{eq:tau_full}
\tau^{-1}(\varepsilon) & =&\frac{\pi Z^2 e^4 n_i}{\varepsilon p^3} \sqrt{\frac{M}{{2 \pi}T }} \!\int_{-\infty}^{\varepsilon-m}\! d \omega\, e^{-\omega / 2 T} \frac{f^0(\varepsilon-\omega)}{f^0(\varepsilon)} \nonumber\\
&\times& \int_{q_{-}}^{q_{+}}\! d q\left(q^2-\omega^2+2 \varepsilon \omega\right) \frac{(2 \varepsilon-\omega)^2-q^2}{\left|q^2+\Pi_L\right|^2} \nonumber\\
&\times & e^{-\omega^2 M/ 2 q^2 T} e^{-q^2 / 8 M T} S(q) F^2(q),
\end{eqnarray}
where $q_{ \pm}=\left| \sqrt{p^2-\left(2 \omega \varepsilon-\omega^2\right)}
  \pm p\right|$, and $\Pi_{L}\simeq 4e^2p_F^2/\pi$ is the longitudinal
component of the polarization tensor. Note that we neglected the
transverse part of the scattering in Eq.~\eqref{eq:tau_full} as that
part is negligibly small in the regime of interest of this work.

If the magnetic field is directed
along the $z$-axis, then the conductivity 
tensor has the form
\bea\label{eq:sigma3}
\hat{\sigma}=
\begin{pmatrix}
    \sigma_0 & -\sigma_1 & 0 \\
    \sigma_1 & \sigma_0 & 0 \\
    0 & 0 & \sigma
\end{pmatrix},
\eea
where the scalar conductivity is given by 
\bea\label{eq:sigma}
\sigma=\sigma_0+\sigma_2=\frac{e^2}{3\pi^2T}\!\int_m^\infty\!
d\varepsilon\, \frac{p^3}{\varepsilon}\tau f^0(1-f^0).
\eea
In the absence of a magnetic field, the conduction becomes isotropic
with $\vecj=\sigma\vecE$, where $\sigma$ is referred below as scalar
conductivity.

At relatively low temperatures $T\leq 5$~MeV the electronic gas in the inner crust is practically in the degenerate state, therefore we can use the low-temperature limit of
Eqs.~\eqref{eq:sigma2} and \eqref{eq:sigma} by substituting
$\partial f^0/\partial\varepsilon=- f^0(1-f^0)/T \to
-\delta(\varepsilon-\varepsilon_F)$, which leads us to the well-known
Drude formulas
\bea\label{eq:sigmas_fermi}
\sigma =\frac{n_ee^{2}\tau}{\varepsilon}\bigg\vert_{\varepsilon=\varepsilon_F},\quad 
\sigma_0=\frac{\sigma}{1+(\omega_{c}\tau)^2}\bigg\vert_{\varepsilon=\varepsilon_F},\quad
\sigma_1= (\omega_{c}\tau)\bigg\vert_{\varepsilon=\varepsilon_F}\!\!\!\sigma_0,
\nonumber\\
\eea
where the quantities $\tau$ and $\omega_c$ in these equations
are evaluated at the Fermi energy $\varepsilon_F$.

\section{Numerical results}
\label{sec:Results}

Numerically, the electrical conductivity is evaluated using the
relaxation time Eq.~\eqref{eq:tau_full}.  With this relaxation time,
we evaluate the components of the conductivity tensor using
Eq.~\eqref{eq:sigma2}.  Two different regimes of weak and strong
magnetic fields arise which are distinguished by the scalar vs. tensor
nature of the conductivity, see Eq.~\eqref{eq:sigmas_fermi}.  These
regimes are distinguished by the value of the {\it Hall parameter}
$\omega_c\tau$.  {Note that as, by definition, $\tau$ and $\omega$ are
  energy-dependent quantities, they should be evaluated at the Fermi
  energy in the degenerate regime, which is the case for
  $T\leq 5$~MeV. At high temperatures $T\geq T_F/3$ where electrons
  are no longer degenerate, the quantities $\tau$ and $\omega$ are
  evaluated at the thermal energy $\bar{\varepsilon}=3T$, see below
  (this does not apply, naturally, to the cases where $\tau$ and
  $\omega$ are integrated over the energies where the
  energy dependence should be kept).}  In the isotropic case
$\omega_c\tau \ll 1$ one has $\sigma_1\ll \sigma_0\simeq\sigma$,
therefore, all three diagonal components of the conductivity tensor
are identical, and the non-diagonal components vanish. In the
anisotropic case $\omega_c\tau \simeq 1$ they are distinct and should
be studied separately.  Below, we will study the dependence of the
conductivity on the density, temperature, and strength of the magnetic
field for the selected compositions.

\subsection{Relaxation time and the Hall parameter}

\begin{figure}[bt] 
\begin{center}
\includegraphics[width=\linewidth,keepaspectratio]{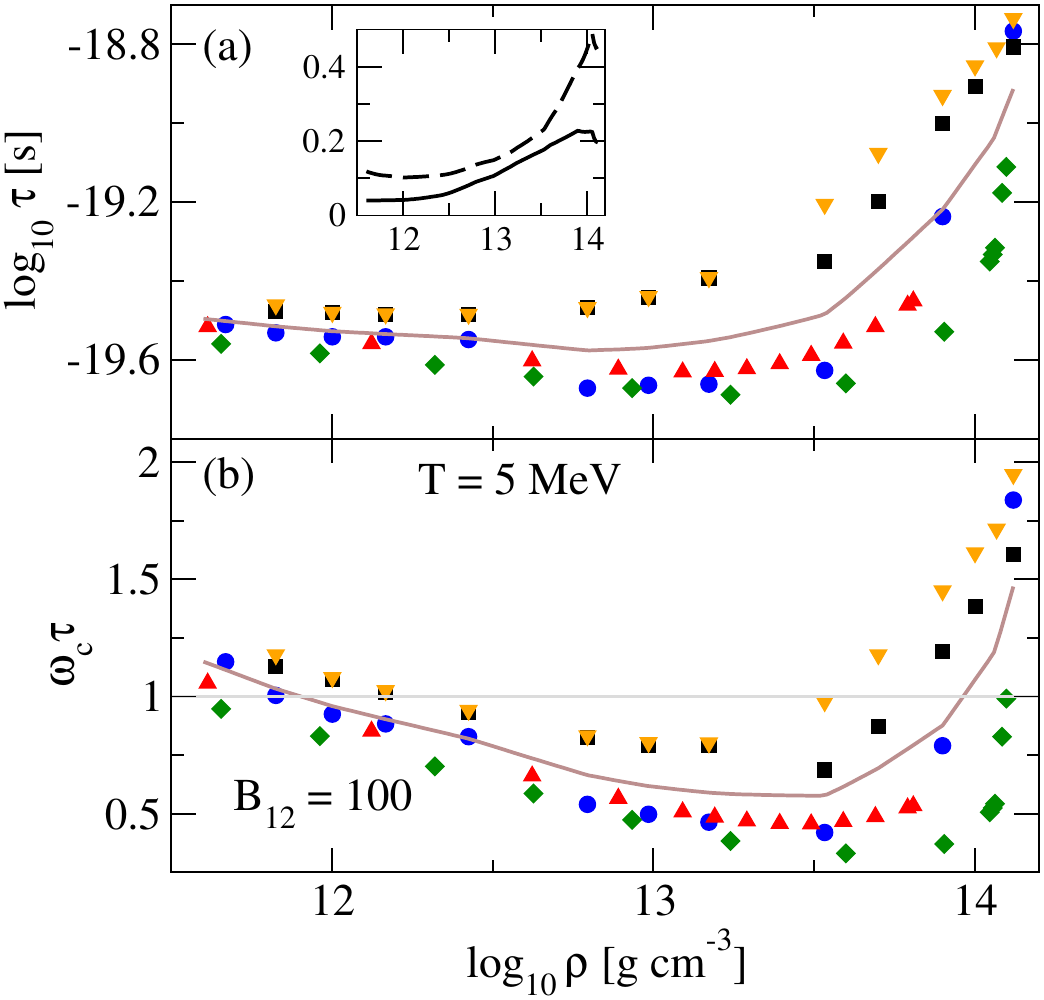}
\caption{ The relaxation time $\tau$ and the Hall parameter
  $\omega_{c}\tau$ at the Fermi energy as functions of the mass
  density for five compositions as labeled in
  Fig.~\ref{fig:compositions}. The temperature is fixed at $T=5$~MeV,
  and the magnetic field is fixed at $B_{12}=100$ for (b). The solid
  lines show the values of these quantities averaged over the five
  compositions. The solid and dashed curves in the inset show the
  standard deviations of $\log_{10}\tau$ and $\omega_{c}\tau$,
  respectively.}
\label{fig:tau_dens}
\end{center}
\end{figure}
\begin{figure}[hbt] 
\begin{center}
\includegraphics[width=\linewidth,keepaspectratio]{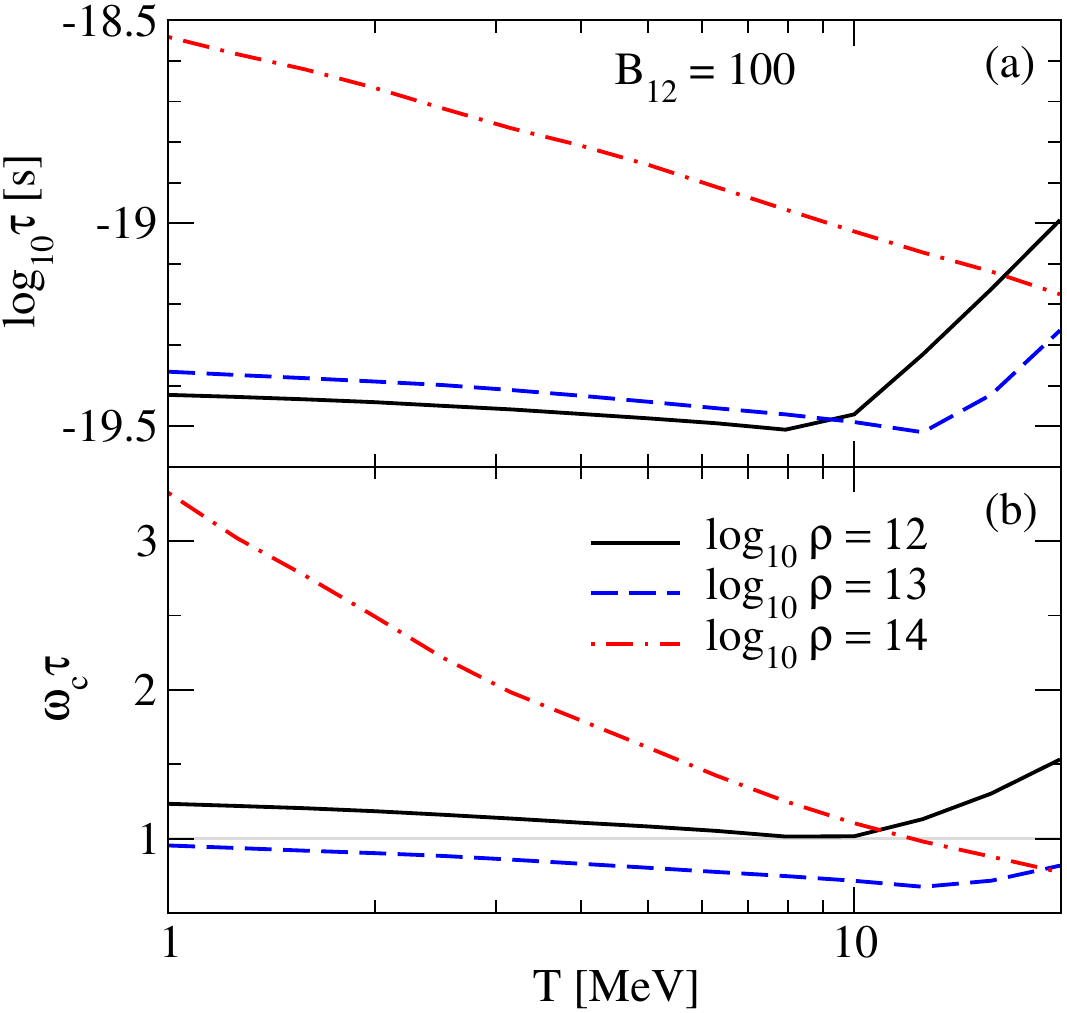}
\caption{ The relaxation time $\tau$ and the Hall parameter
  $\omega_{c}\tau$ as functions of the temperature for various values
  of the mass density for composition D1M$^*$ ($\tau$ and $\omega_{c}$
  are evaluated at $\varepsilon=\varepsilon_F$ if $T\leq T_F/3$ and at
  $\varepsilon=3T$ if $T\geq T_F/3$, see the discussion in the
  text). In panel (b) the magnetic field is fixed at $B_{12}=100$.}
\label{fig:tau_temp}
\end{center}
\end{figure}

Figure~\ref{fig:tau_dens} shows the relaxation time $\tau$ and the Hall parameter $\omega_c\tau$ for five compositions as functions of the mass density for the temperature  $T= 5~{\rm MeV}$. Because it corresponds to the degenerate regime for electrons,  $\tau$ and $\omega_c$ are evaluated at the Fermi energy. To assess the variations in the relaxation time and Hall parameter $\omega_c\tau$ with the composition  we use the same approximations 
as in Ref.~\cite{Harutyunyan2016}, to reduce Eq.~\eqref{eq:tau_full} to the following  form first obtained  by Ref.~\cite{Nandkumar1984MNRAS}
\bea\label{eq:tau_approx}
\tau^{-1}=\frac{4 Z e^4 \varepsilon_F}{3 \pi}\! \int_0^{2 p_F}\! \frac{d q}{q}\left(1-\frac{q^2}{4 \varepsilon_F^2}\right) S(q)F^2(q),
\eea
which implies the scaling 
\bea \label{eq:tau_scaling}
\tau\sim Z^{-4/3} n_i^{-1/3},
\eea
if we neglect the effects of $S(q)$ and $F(q)$, where we used
$\varepsilon_F = p_F \sim (Zn_i)^{1/3} $. Because the value of $Z$ for
most of the compositions is fixed at (semi-)magic number 40 or 50,
Eq.~\eqref{eq:tau_scaling} suggests decreasing relaxation times with
the density as follows from Fig.~\ref{fig:neutrons}. However, such
behavior is realized only in the low-density domain $\log_{10}\rho$
[g~cm$^{-3}]\le 12.5\div 13$. At higher densities
$\log_{10}\rho\, [\textrm{g\, cm}^{-3}] >13$ the relaxation time
reverses from slowly decreasing function to an increasing function of
density, which is a consequence of the increasing importance of finite
size of the nuclei encoded in the nuclear form factor. It suppresses
the electron-ion scattering rates significantly at high densities
where the nuclear radii become close to the radii of the Wigner-Seitz
cell. This results in larger relaxation times at higher densities.
  
  This effect is stronger for two models D1M and D1M$^*$ as these
  models predict a significant increase in the mass number and size of
  nuclei with the density in the high-density regime, as seen from
  Fig.~\ref{fig:compositions}.  Figure~\ref{fig:tau_sigma_noform},
  panel (a) shows the relaxation time for two models D1M$^*$ and
  Bsk24, which predict the same values of $Z$ and very close values of
  $n_i$, therefore, the differences in the relaxation times predicted
  by these models can be attributed to the difference in the values of
  $A$, which manifests itself through the nuclear form factor. Indeed,
  as seen from the figure, the relaxation times for both models are
  very similar and do follow the scaling $\tau\sim n_i^{-1/3}$ in the
  case were the factor $F(q)$ is set to one (empty symbols) whereas in
  the full calculation (filled symbols) $\tau$ increases with the
  density with different slopes for different models, thus the
  scaling~\eqref{eq:tau_scaling} fails.  The
  scaling~\eqref{eq:tau_scaling} works qualitatively well for outer
  crust matter, where the effect of the form factor is small for
  nuclei with small $A$~\cite{Harutyunyan2016}.

  The solid line in Fig.~\ref{fig:tau_dens}(a) shows the average
  logarithmic relaxation timescale computed as
  $\langle\log_{10}\tau\rangle=\sum_{i=1}^5\log_{10}\tau_i/5$, where
  $\tau_i$ are the relaxation times for the five compositions
  interpolated in the density range
  $11.6\leq\log_{10}\rho\, [\textrm{g\, cm}^{-3}] \leq 14.12$.
  Similarly, the solid line in Fig.~\ref{fig:tau_dens}(b) shows the
  average value of $\omega_c\tau$.  In addition, we show the standard
  deviations of these quantities from their average values in the
  inset located in  Fig.~\ref{fig:tau_dens}(a).  The solid
  line shows the standard deviation for $\log_{10}\tau$, and the
  dashed line -- that for $\omega_c\tau$.  We compute the standard
  deviation $s_a$ of a quantity $a$ using the formula
  $s_{a}=\sqrt{\sum_{i=1}^5(a_i-\langle a \rangle)^2/5}$, where
  $\langle a \rangle$ is the average value over the five compositions.
\begin{figure}[bt] 
\begin{center}
\includegraphics[width=\linewidth,keepaspectratio]{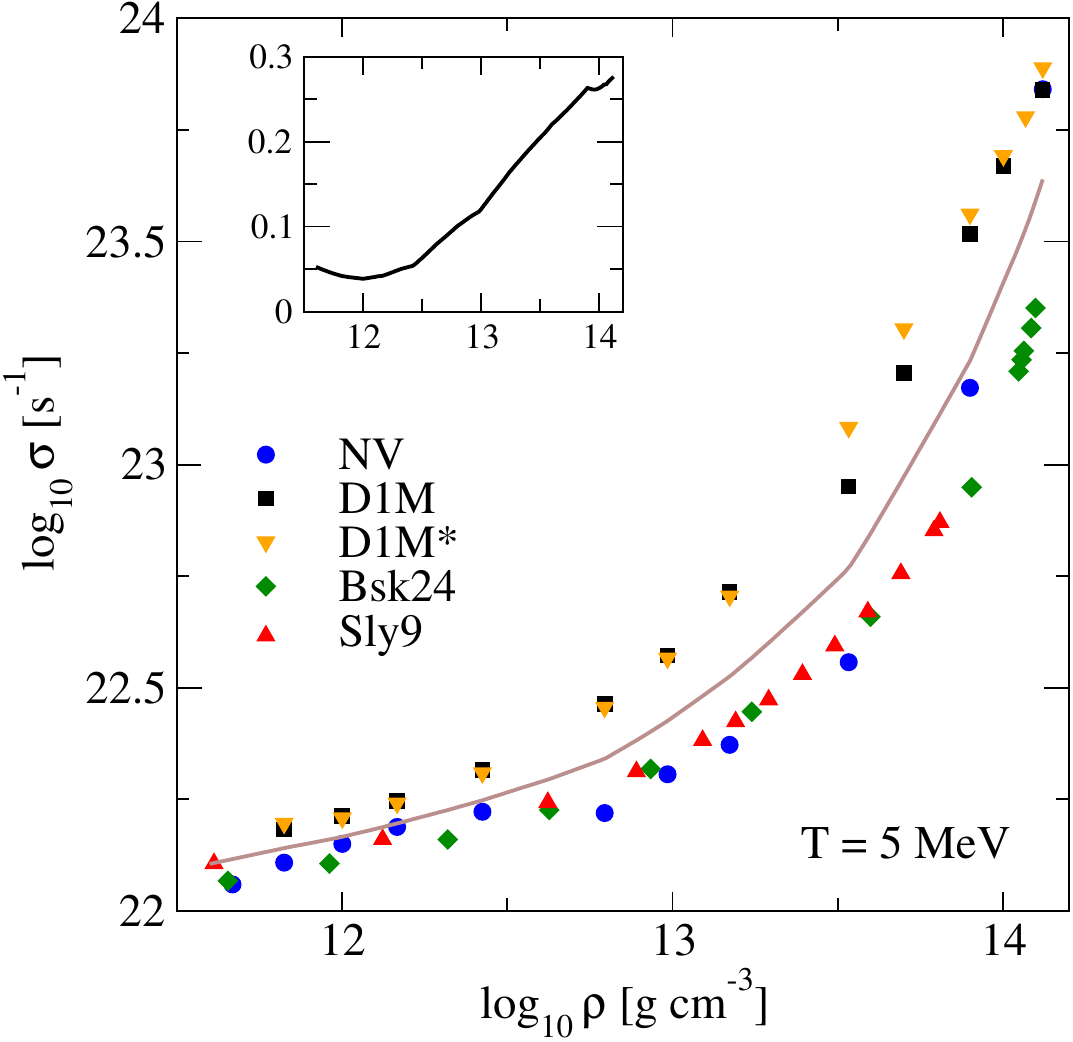}
\caption{Dependence of the scalar conductivity on density for five
  compositions. The temperature is fixed at $T=5$~MeV.  The solid line
  shows the logarithm of the conductivity averaged over the five
  compositions, and the inset shows the standard deviation for
  $\log_{10}\sigma$.}
\label{fig:sigma_dens}
\end{center}
\end{figure}
\begin{figure}[bt] 
\begin{center}
\includegraphics[width=\linewidth,keepaspectratio]{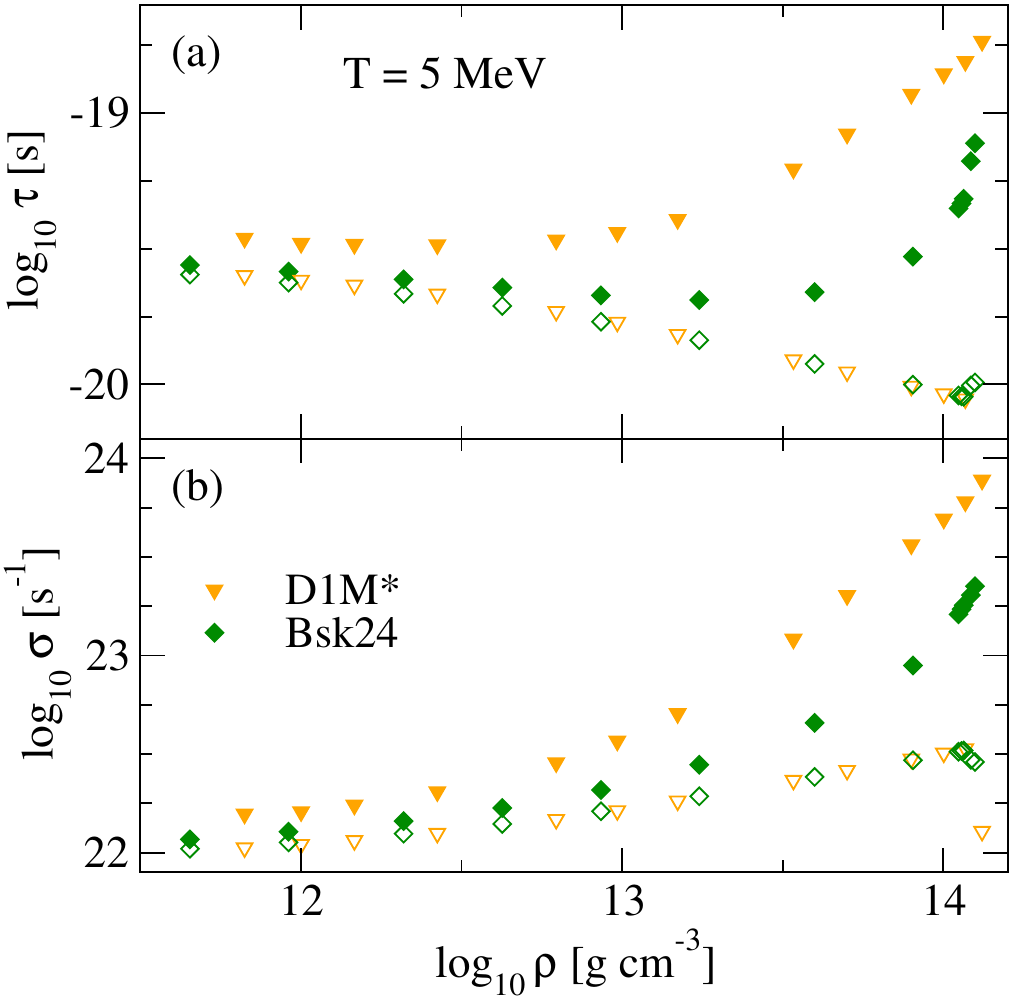}
\caption{(a) Dependence of the relaxation time and (b) the scalar
  conductivity on density for two models evaluated with full
  nuclear form factor $F(q)$ (filled symbols) and with $F(q)=1$ (empty
  symbols). The temperature is fixed at $T=5$~MeV.}
\label{fig:tau_sigma_noform}
\end{center}
\end{figure}
\begin{figure}[hbt] 
\begin{center}
\includegraphics[width=\linewidth,keepaspectratio]{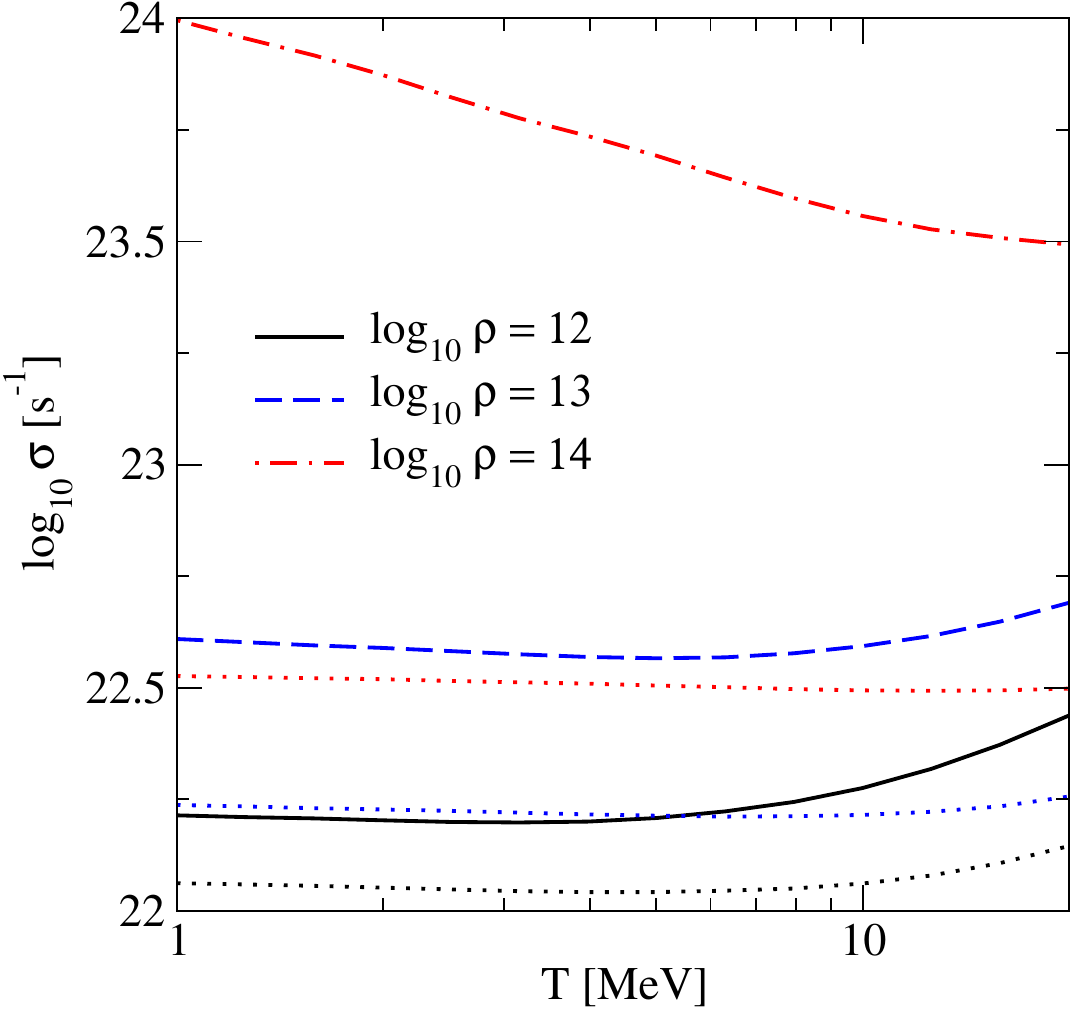}
\caption{ The temperature dependence of the scalar 
conductivity for various values of the density for composition D1M$^*$.
The dotted lines show the corresponding values of $\sigma$
in the case where the effect of the nuclear form factor is neglected.}
\label{fig:sigma_temp}
\end{center}
\end{figure}
  
Figure~\ref{fig:tau_temp} shows the temperature dependence of the
relaxation time and the Hall parameter for several densities for the
composition D1M$^*$.  Here we extrapolate our results of
low-temperature matter to higher temperatures (up to 20~MeV) making a
crude assumption that the composition of matter does not change
significantly with the temperature at the given density.  As the
matter is partially degenerate at temperatures $T\geq 10$~MeV at low
densities, we evaluate the quantities $\tau$ and $\omega_c$ at the
Fermi energy at temperatures $T\leq T^*$, and at the thermal energy of
ultrarelativistic electrons $\bar\varepsilon = 3T$ at $T\geq T^*$,
where $T^*=T_F/3$ is the transition temperature from the
non-degenerate to the degenerate regime and corresponds to the
requirement that the Fermi energy becomes equal to the thermal energy
of a nondegenerate gas, \ie,
$\varepsilon_F\simeq T_F=3T$~\cite{Harutyunyan2016}. It is seen that
$\tau$ decreases with the temperature in the strongly degenerate
regime $T\le T^*$ and increases in the semi-degenerate regime
$T\ge T^*$, which leads to a minimum at $T\sim T^*$ in the
conductivity. The first effect originates from the structure factor
$S(q)$ and, at very high densities also from the nuclear form factor
$F(q)$.  In the semi-degenerate regime, the temperature dependence of
$\tau $ is dominated by the energy increase of electrons with
temperature, and $\tau$ becomes increasing with temperature.

  From the bottom panels of Figs.~\ref{fig:tau_dens} and
  \ref{fig:tau_temp} we see that for magnetic field
  $B_{12}\equiv B/(10^{12}{\rm G})=100$ the factor $\omega_c\tau$ is
  of the order of unity in the whole inner crust. This implies that
  the effect of anisotropy should become important at such values of
  fields. We see that the effects of anisotropy in the inner crust are
  less pronounced than in the low-density outer crust, where the
  anisotropy becomes important already for
  $B_{12}\geq 0.01$~\cite{Harutyunyan2016}.

\subsection{Conductivity in the low-field limit}

We start with the results on the density dependence of the scalar
conductivity at a fixed temperature.  Figure~\ref{fig:sigma_dens}
shows the scalar conductivity as a function of the density for
$T=5$~MeV. Despite the non-monotonic behavior of the relaxation time
with the density, the increase of density of the states close to the
Fermi surface leads to an increase of conductivity with matter
density, as seen from the first formula of
Eq.~\eqref{eq:sigmas_fermi}.  Similar to the case of the relaxation
time, comparisons can be made among the various compositions and the
effects of various factors on the conductivity.  Given that for most
of the compositions $Z$ is fixed at a (semi-)magic number, and the ion
number density $n_i$ has similar values, the dependence of $\sigma$ on
the density for any given composition is controlled mainly by the
dependence of the values of $A$ on density, which affects the
conductivities by means of the nuclear form factors, as mentioned
above.  Fig.~\ref{fig:tau_sigma_noform}(b) shows the conductivity for
two models D1M$^*$ and Bsk24 and two cases with $F(q)=1$ and full
$F(q)$ which confirm the statements above. We see that for $F(q)=1$
the scalar conductivity would increase as a power low with the density
following the universal approximate scaling $\sigma\propto\rho^{1/4}$,
whereas for full $F(q)$ a much faster increase of the conductivity at
high densities is observed.  The slopes are composition-dependent.

In addition, we show in Fig.~\ref{fig:sigma_dens} the (logarithmic)
conductivity averaged over the five compositions and the standard
deviation for $\log_{10}\sigma$ in the inset. As expected, the
standard deviation rises with density as the differences between the
compositions with regard to predicted values of $Z$, $A$, and $n_i$
increase with density. The deviations rise sharply beyond
$\log_{10}\rho\, [\textrm{g\, cm}^{-3}] \geq 13$, but the overall
deviation remains below $25\%$, except close to the crust-core
interface, where non-spherical nuclei are expected.

The temperature dependence of the conductivity at fixed values of the
density is shown in Fig.~\ref{fig:sigma_temp}.  This figure allows for
extrapolation to larger temperatures where the compositions of the
matter used are not realistic as they have been derived at zero
temperature and assuming matter without neutrinos. In the range of
temperatures where the compositions are realistic $0\le T\le 5$~MeV
(and beyond up to 10 MeV) the temperature dependence of the
conductivity is very weak, except at very high densities and models
with very large values of $A$, where the nuclear form factor plays the
key role in the scattering rates as discussed in the previous
subsection. To illustrate this, we show in Fig.~\ref{fig:sigma_temp}
the corresponding values of $\sigma$ in the case where the effect of
nuclear form factor is neglected, \ie, $F(q)=1$. We clearly see that
the form factor affects not only the magnitude of the conductivity at
high densities, but also modifies its temperature dependence by making
it much steeper than at low density. Numerically we find that for the
model D1M$^*$ the form factor increases the conductivity by factors of
1.4, 2.3, and between $15$ and $30$ at densities $\rho=10^{12}$, $\rho=10^{13}$,
and $\rho=10^{14}$~g~cm$^{-3}$, respectively, and for
$5\geq T\geq 1$~MeV.  These factors are similar for the model D1M
which predicts very large values of $A$ at
$\rho\ge 10^{13}$~g~cm$^{-3}$ as well. The remaining models are
characterized by smaller values of $A$, therefore the modifications
due to the nuclear form factor are less pronounced for
$\rho\simeq 10^{14}$~g~cm$^{-3}$.

\begin{figure}[hbt] 
\begin{center}
\includegraphics[width=\linewidth,keepaspectratio]{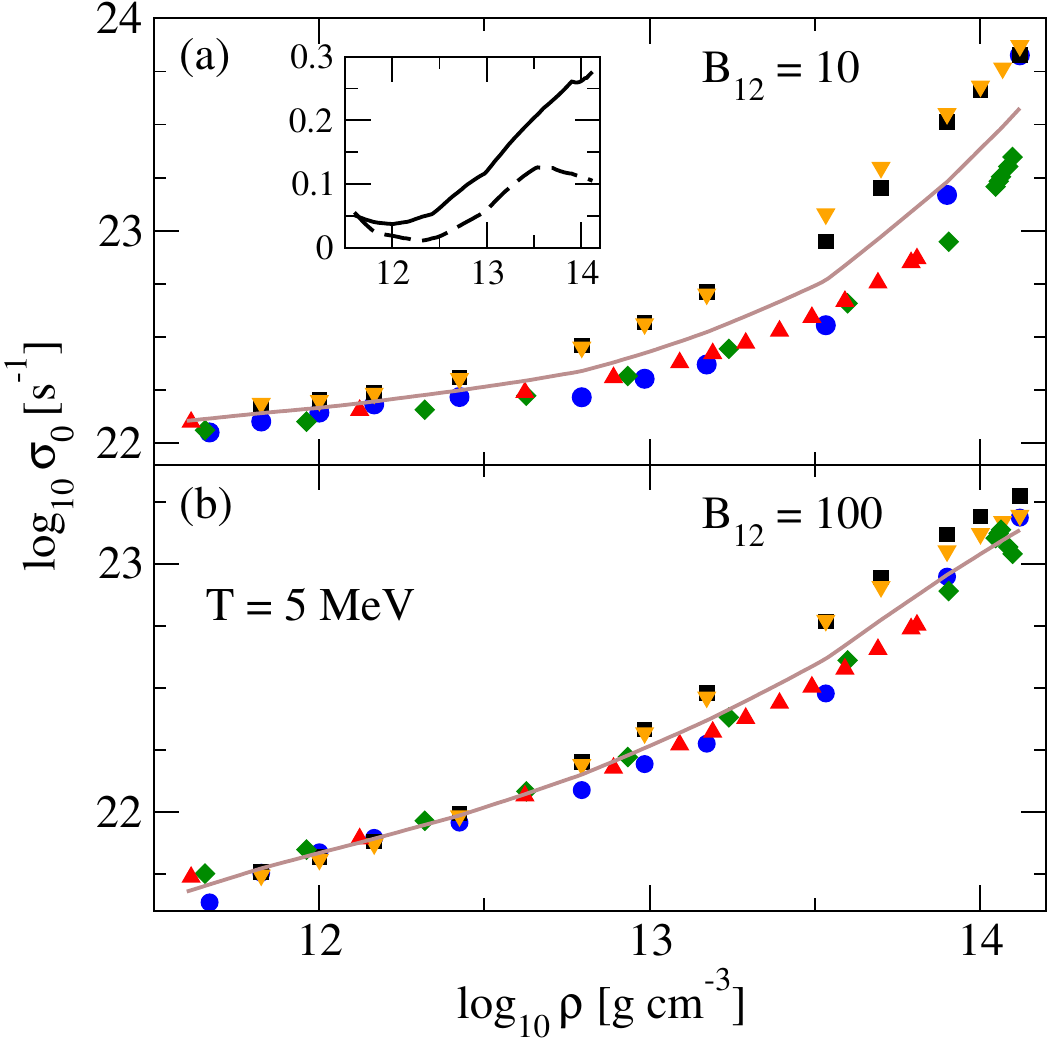}
\caption{ Dependence of $\sigma_0$ component of the electrical
  conductivity tensor on density for five compositions. The labeling
  of the curves is as in Fig.~\ref{fig:sigma_dens}.  The values of the
  temperature and the magnetic field are indicated in the plot.  The
  solid lines in each panel show the averages of $\log_{10}\sigma_0$
  over the five compositions. The solid and dashed curves in the inset
  show the standard deviations of $\log_{10}\sigma_0$ for $B_{12}=10$
  and $B_{12}=100$, respectively.}
\label{fig:sigma0_dens}
\end{center}
\end{figure}
\begin{figure}[hbt] 
\begin{center}
\includegraphics[width=\linewidth,keepaspectratio]{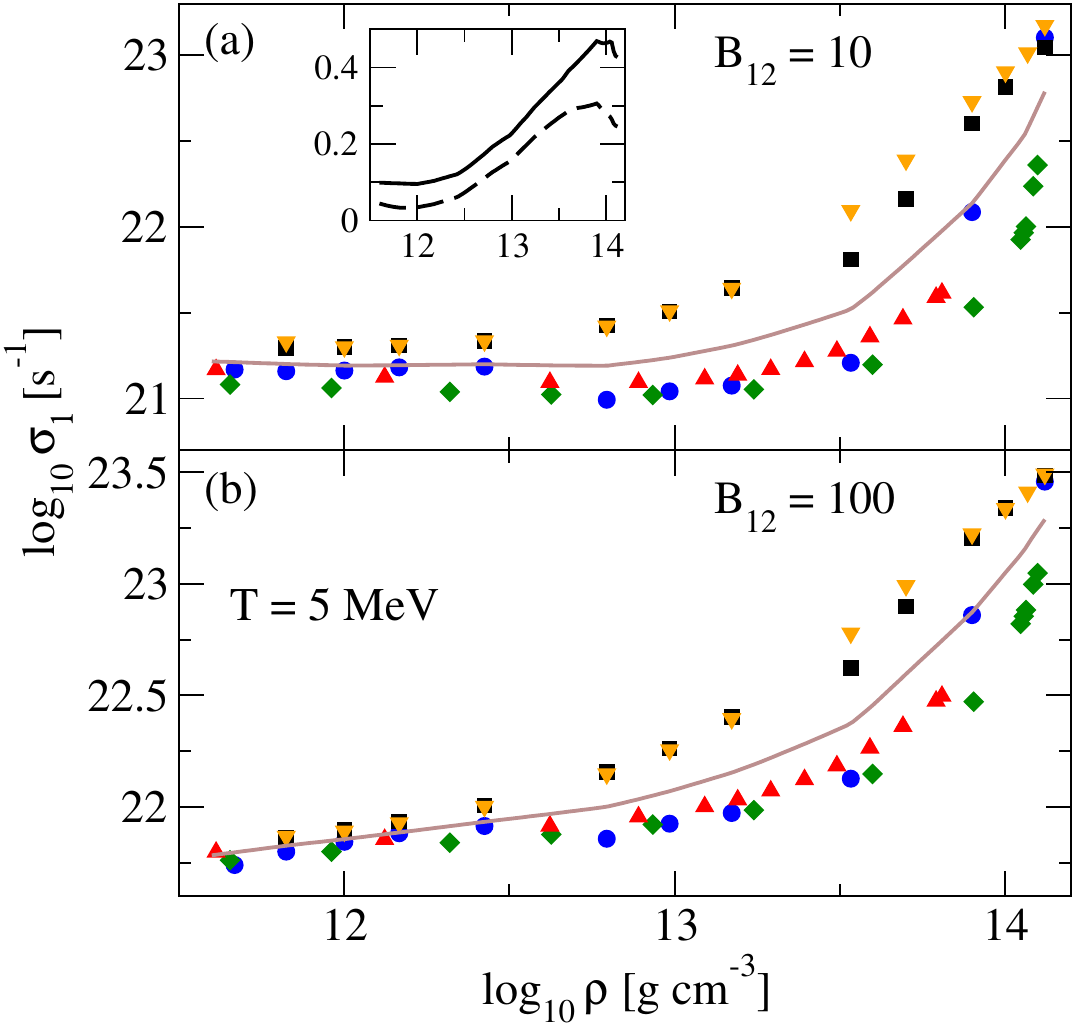}
\caption{ Dependence of $\sigma_1$ component of the electrical
  conductivity tensor on density for four compositions. The labeling
  of the curves is as in Fig.~\ref{fig:sigma_dens}. The values of
  the temperature and the magnetic field are indicated in the plot. 
  The solid lines in each panel show the averages of $\log_{10}\sigma_1$ 
  over the five compositions. The solid and dashed curves in 
  the inset show the standard deviations of $\log_{10}\sigma_1$ 
  for $B_{12}=10$ and $B_{12}=100$, respectively.}
\label{fig:sigma1_dens}
\end{center}
\end{figure}
\begin{figure}[hbt] 
\begin{center}
\includegraphics[width=\linewidth, keepaspectratio]{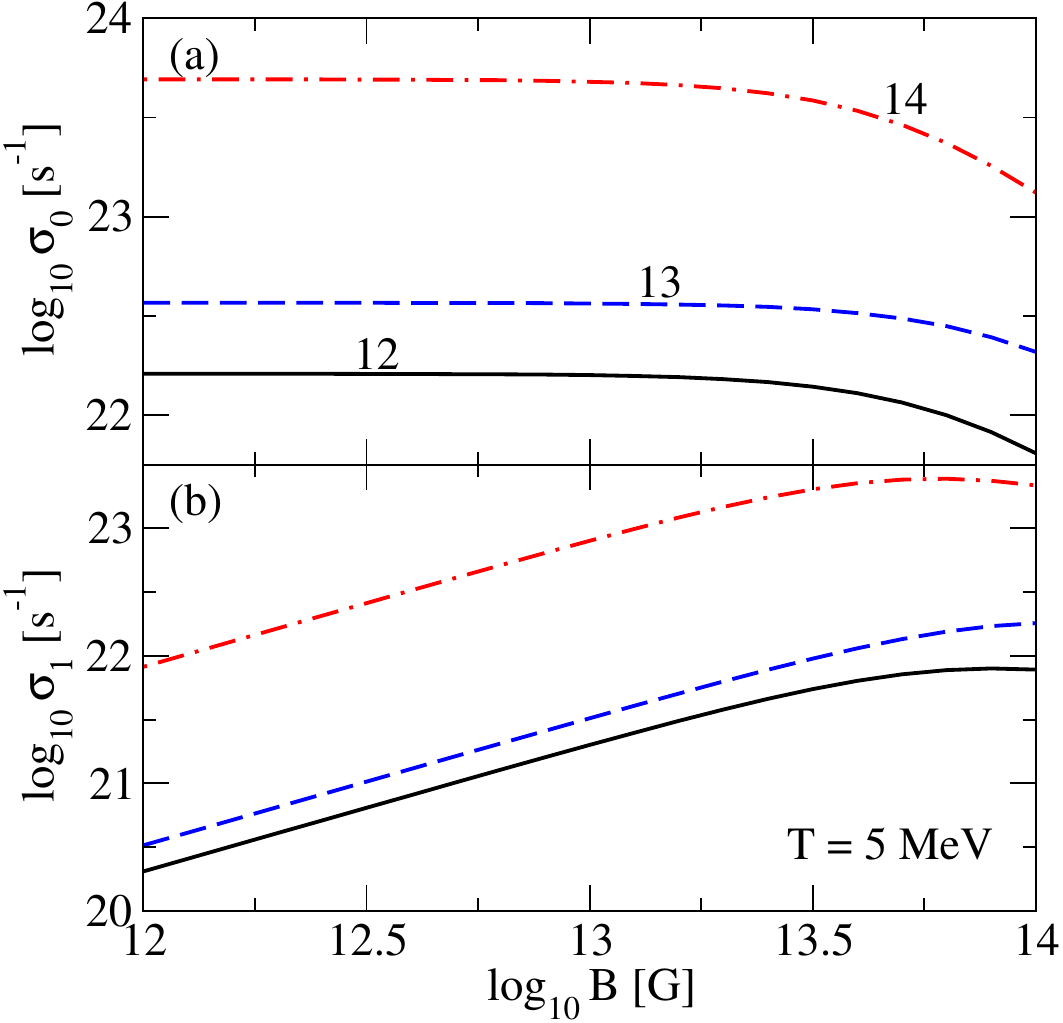}
\caption{ The dependence of $\sigma_0$ and
  $\sigma_1$ components of the electrical
  conductivity tensor on the magnetic field 
  at fixed temperature and for various values of 
  the density indicated on the plot by their logarithm for composition D1M$^*$. }
\label{fig:sigma_b}
\end{center}
\end{figure}
\begin{figure}[!] 
\begin{center}
\includegraphics[width=\linewidth, keepaspectratio]{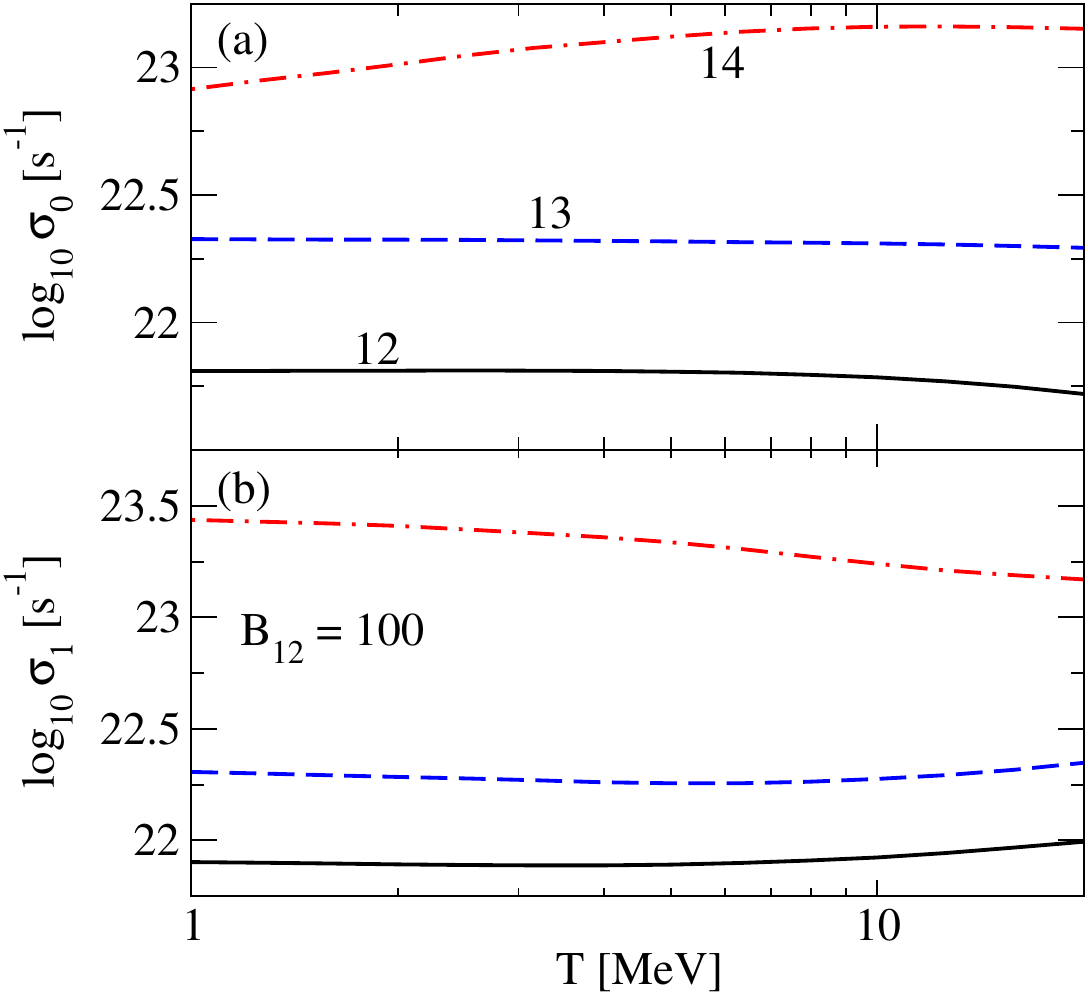}
\caption{ The dependence of $\sigma_0$ and $\sigma_1$ components of
  the electrical conductivity tensor on the temperature at fixed
  magnetic field for various values of the density indicated on the
  plot by their logarithm for composition D1M$^*$.}
\label{fig:sigma01_temp}
\end{center}
\end{figure}

\subsection{Conductivities in the high-field limit}

We now consider the density dependence of the conductivity in the
anisotropic case where the magnetic fields are
large. Figure~\ref{fig:sigma0_dens} shows the density dependence of
the $\sigma_0$ component for two
values of the magnetic field ($B_{12}=10,\,
100$) for compositions considered. The temperature is fixed at $T=5$~MeV.
The same for the component $\sigma_1$ is shown in Fig.~\ref{fig:sigma1_dens}.

For the magnetic field $B_{12}=10$ we have $\omega_c\tau \ll 1$,
which implies essentially isotropic conduction. In this case, the
component $\sigma_0$ is almost identical to the scalar conductivity
$\sigma$ as seen from Figs.~\ref{fig:sigma_dens} and
\ref{fig:sigma0_dens}, and the values of $\sigma_1$ are much lower.
For $B_{12}\gtrsim 30$ the anisotropy already sets in, as can be
seen from Fig.~\ref{fig:sigma_b}, and for the value $B_{12}=100$ the
transverse components of the conductivity $\sigma_0$ and $\sigma_1$
have the same order of magnitude, as already $\omega_c\tau\sim 1$.
The averages of $\log_{10}\sigma_0$ and $\log_{10}\sigma_1$ and
their standard deviations are also shown in Figs.~\ref{fig:sigma0_dens} 
and \ref{fig:sigma1_dens}.

We show the magnetic field dependence of the components of
conductivities for fixed values of temperature and density in
Fig.~\ref{fig:sigma_b}.  According to Eq.~\eqref{eq:sigmas_fermi}
$\sigma_1$ is proportional to the magnetic field in the isotropic
regime whereas $\sigma_0$ is independent of it; these features are
seen in Fig.~\ref{fig:sigma_b}.  In general, we see that the effect of
the magnetic field on the conduction in the inner crust up to the limit of
non-quantizing fields $B_{12}\simeq 100$ is not as significant as it
was the case for the outer crust, which becomes totally anisotropic for
$B_{12}\geq 10$~\cite{Harutyunyan2016}.

The temperature dependence of the conductivity components $\sigma_0$
and $\sigma_1$ is shown in Fig.~\ref{fig:sigma01_temp} for fixed
magnetic field $B_{12}=100$. These components
depend weakly on the temperature,
because the temperature dependence of the scalar conductivity $\sigma$ and that of the product
$\omega_c\tau$ partially cancel each other.

\subsection{Remarks on electron-neutron scattering}
{
In addition to the Coulomb scattering of electrons off the nuclei 
studied in detail above, electrons will be scattered by neutrons due 
to their anomalous magnetic moment
$g_n=-1.91$. The ratio of the respective relaxation times can be 
estimated from the transition probabilities $W_{ep}$ and $W_{en}$ 
for electron-proton and electron-neutron scattering which were 
computed in Ref.~\cite{Flowers:1979} and are given by their 
Eqs.~(48) and (49). The ratio of the corresponding relaxation 
times, up to a factor $O(1)$, is given by 
\bea\label{eq:ratio}
\frac{\tau_{en}}{\tau_{ep}} \sim \frac{W_{ep} Z}{W_{en} N} \simeq
\frac{1}{2g_n^2} \left(\frac{m_N}{p_F}\right)^2 \frac{Z}{N},
\eea
where $m_N$ is the neutron mass and $N$ is the number of neutrons in a Wigner-Seitz cell. For an estimate, we use the model D1M and Table II of Ref.~\cite{Mondal2020} in the range  $0.006\ge n_B\ge 0.06$  [fm$^{-3}]$
(which corresponds to the mass density range $12\le \log_{10}\rho$ [g~cm$^{-3}]\le 14$). We then find that
$0.185\ge {Z}/{N}\ge 0.04$ and $1175 \ge (m_N/p_F)^2 \ge 138$, assuming bare neutron mass $m_N=939$~MeV, and 
\bea
30 \ge \frac{\tau_{en}}{\tau_{ep}}\ge 0.8.
\eea
From this elementary estimate, we conclude that the electron-neutron interaction can appreciably contribute to the conductivity close to the crust-core transition, but is subdominant at lower densities. 
}

\section{Conclusions}
\label{sec:Conclusions}

In this work, we extended the study of the finite-temperature
conductivity of the outer crust of compact stars in the liquid regime
in magnetic fields \cite{Harutyunyan2016} to the inner crust
characterized by a neutron drip component that forms a separate fluid.
Because of higher densities, the electrons in the inner crust are
mostly in the degenerate regime within the range of temperatures in
which the composition of the inner crust computed at zero temperature
can still be applied ($T\le T_{\rm tr}\simeq 5$~MeV). Nevertheless, our extrapolation
beyond this range shows that there is a minimum in the conductivity at
the transition from the degenerate to the non-degenerate regime, which
was also the case in the outer crust.  
Of course, a self-consistent computation of the conductivity
of multi-component nuclear plasma in this regime is required to draw
definitive conclusions.

To assess the uncertainties in the conductivities due to the
composition we have taken an average over the five adopted
compositions (NV, Bsk24, D1M, D1M$^*$, and Sly9) and computed the
standard deviation. The deviations are below $10\%$ for densities
$\log_{10} \rho\, [\textrm{g\, cm}^{-3}]\leq 12.5$ and increase up to
$25\%$ for higher densities.  Exceptions are very high densities
$\log_{10}\rho\,[\textrm{g\, cm}^{-3}] \ge 14$ where standard
deviations can increase up to $40\%$, but one anticipates
non-spherical nuclei at such densities. Our analysis shows that this
scatter among different compositions affects the transport quantities
through the nuclear form-factor (i.e., finite nuclear size).  It is
the dominant agent that determines the behavior of the conductivity at
high densities.

The anisotropy of transport becomes sizable for magnetic fields
$B_{12}\ge 30$ and low densities as well as for larger fields $B_{12}\ge 50$
at higher densities. The field diminishes the $\sigma_{0}$ component of
the conductivity and the non-diagonal component $\sigma_1$ becomes of
the order of $\sigma_0$ at $B_{12}\simeq 100$.

{ Dissipation through conductivity may or may not be important
  depending on the effectiveness of the other channels of dissipation
  appearing in the relativistic fluid dynamics description of BNS
  mergers, supernovas, and proto-neutron stars. Recent microscopic
  work shows that the bulk viscosity is a potentially important
  channel of
  dissipation~\cite{Schwenzer2018PhRv,Alford2019a,Alford2019b,*Alford2021a,*Alford2020a,*Alford2023}
  as can be seen from the estimates of relevant damping
  timescales~\cite{Alford2021b,Alford2022Particles} and
  implementations in the relativistic hydrodynamics
  simulations~\cite{Most2022,Celora2022}.  }

{ Looking ahead, it would be interesting to apply the formalism used in this work to study the conductivity of the star's core. At low temperatures, the core is superconducting, and conductivity and screening are affected by heterogeneieties on the scales of the order of $10^{2}-10^{4}$~fm, which are introduced by the flux-tubes~\cite{Kobyakov:2023} in the case of type-II and normal domains in the case of type-I superconductivity~\cite{Sedrakian:2005}. 
}

To conclude, we have quantified the conductivity of the warm inner crust
of a compact star within the Boltzmann quasiparticle transport of
electrons in the liquid phase of inner crust matter taking into account the screening of
electron-ion interaction, finite nuclear size, ion structure factor,
and magnetic fields. Five different compositions were studied to assess
the dependence of the results on adopted composition.  In the future, this study can be extended to higher temperatures by adopting
compositions that have been derived at finite temperature and account
for the multi-ion composition of matter, which includes  $\alpha$-particles
and other light clusters.

\section*{Acknowledgements}
The authors gratefully acknowledge the collaborative research grant No. 96 839
of Volkswagen Foundation (Hannover, Germany). We thank  Jonas Dittrich 
for testing the code used in this work and N.\,Chamel,  C.\,Mondal, and A.\,Raduta for useful communications.
A.\,S.  acknowledges
support through  Deutsche Forschungsgemeinschaft Grant No. SE
1836/5-2, the Polish NCN Grant No. 2020/37/B/ST9/01937,
and thanks the Institute for Nuclear Theory at the University
of Washington for its kind hospitality, stimulating research 
environment, and partial support by the INT's U.S. Department of 
Energy grant No. DE-FG02-00ER41132.

\bibliographystyle{apsrev}     
\bibliography{BibliographyCrustTrans}

\newpage
\begin{widetext}
\appendix

\section{Tables of transport parameters}

In this appendix we provide numerical results for the density
dependence of the components of conductivity tensor along with
relaxation time and anisotropy parameter for the five models studied
in this work. Tables are given for fixed temperatures $T=1$ and $T=5$
MeV, which bracket the range to which the current study is applicable
and fixed value of magnetic field $B_{12}=100$. Table 1 is computed
with  model NV, Table 2 with  model D1M, Table 3 with model D1M*,
Table 4 with  model Bsk24, and Table 5 with  model Sly9.

\begin{table*}[b]
\begin{center}
\caption {Model NV and $B_{12}=100$. } \label{tab:sigma_1}
\vskip 0.2cm
{\scriptsize
\begin{tabular}{ c | c c c c c | c c c c c }
\hline
&  &  &  $T$ = 1 MeV  &  &  &  &  &  $T$ = 5 MeV  &  &  \\ 
\hline
$\log_{10}\rho$ & $\log_{10}\tau$ & $\omega_c\tau$ & $\log_{10}\sigma$ & $\log_{10}\sigma_0$ & $\log_{10}\sigma_1$ & $\log_{10}\tau$ & $\omega_c\tau$ & $\log_{10}\sigma$ & $\log_{10}\sigma_0$ & $\log_{10}\sigma_1$ \\ 
\hline
\hspace{0.05cm} 11.669 \hspace{0.05cm}  &  \hspace{0.05cm} -19.461 \hspace{0.3cm}  &  1.289  &  22.062  &  21.635 \hspace{0.3cm}  &  21.747 \hspace{0.05cm}  &  \hspace{0.05cm} -19.511 \hspace{0.3cm}  &  1.148  &  22.059  &  21.634 \hspace{0.3cm}  &  21.740 \hspace{0.05cm}  \\ 
\hspace{0.05cm} 11.825 \hspace{0.05cm}  &  \hspace{0.05cm} -19.482 \hspace{0.3cm}  &  1.128  &  22.115  &  21.757 \hspace{0.3cm}  &  21.810 \hspace{0.05cm}  &  \hspace{0.05cm} -19.532 \hspace{0.3cm}  &  1.005  &  22.108  &  21.754 \hspace{0.3cm}  &  21.800 \hspace{0.05cm}  \\ 
\hspace{0.05cm} 12.001 \hspace{0.05cm}  &  \hspace{0.05cm} -19.491 \hspace{0.3cm}  &  1.039  &  22.158  &  21.839 \hspace{0.3cm}  &  21.857 \hspace{0.05cm}  &  \hspace{0.05cm} -19.542 \hspace{0.3cm}  &  0.925  &  22.150  &  21.836 \hspace{0.3cm}  &  21.844 \hspace{0.05cm}  \\ 
\hspace{0.05cm} 12.167 \hspace{0.05cm}  &  \hspace{0.05cm} -19.491 \hspace{0.3cm}  &  0.994  &  22.198  &  21.898 \hspace{0.3cm}  &  21.897 \hspace{0.05cm}  &  \hspace{0.05cm} -19.542 \hspace{0.3cm}  &  0.883  &  22.188  &  21.894 \hspace{0.3cm}  &  21.882 \hspace{0.05cm}  \\ 
\hspace{0.05cm} 12.425 \hspace{0.05cm}  &  \hspace{0.05cm} -19.497 \hspace{0.3cm}  &  0.934  &  22.234  &  21.961 \hspace{0.3cm}  &  21.932 \hspace{0.05cm}  &  \hspace{0.05cm} -19.549 \hspace{0.3cm}  &  0.829  &  22.222  &  21.956 \hspace{0.3cm}  &  21.915 \hspace{0.05cm}  \\ 
\hspace{0.05cm} 12.795 \hspace{0.05cm}  &  \hspace{0.05cm} -19.632 \hspace{0.3cm}  &  0.592  &  22.224  &  22.093 \hspace{0.3cm}  &  21.867 \hspace{0.05cm}  &  \hspace{0.05cm} -19.672 \hspace{0.3cm}  &  0.540  &  22.219  &  22.087 \hspace{0.3cm}  &  21.858 \hspace{0.05cm}  \\ 
\hspace{0.05cm} 12.985 \hspace{0.05cm}  &  \hspace{0.05cm} -19.624 \hspace{0.3cm}  &  0.547  &  22.316  &  22.202 \hspace{0.3cm}  &  21.941 \hspace{0.05cm}  &  \hspace{0.05cm} -19.665 \hspace{0.3cm}  &  0.498  &  22.306  &  22.192 \hspace{0.3cm}  &  21.925 \hspace{0.05cm}  \\ 
\hspace{0.05cm} 13.173 \hspace{0.05cm}  &  \hspace{0.05cm} -19.620 \hspace{0.3cm}  &  0.512  &  22.387  &  22.285 \hspace{0.3cm}  &  21.996 \hspace{0.05cm}  &  \hspace{0.05cm} -19.662 \hspace{0.3cm}  &  0.464  &  22.372  &  22.274 \hspace{0.3cm}  &  21.973 \hspace{0.05cm}  \\ 
\hspace{0.05cm} 13.533 \hspace{0.05cm}  &  \hspace{0.05cm} -19.579 \hspace{0.3cm}  &  0.469  &  22.584  &  22.497 \hspace{0.3cm}  &  22.169 \hspace{0.05cm}  &  \hspace{0.05cm} -19.627 \hspace{0.3cm}  &  0.420  &  22.557  &  22.477 \hspace{0.3cm}  &  22.127 \hspace{0.05cm}  \\ 
\hspace{0.05cm} 13.900 \hspace{0.05cm}  &  \hspace{0.05cm} -19.105 \hspace{0.3cm}  &  1.070  &  23.290  &  22.959 \hspace{0.3cm}  &  22.988 \hspace{0.05cm}  &  \hspace{0.05cm} -19.237 \hspace{0.3cm}  &  0.790  &  23.172  &  22.948 \hspace{0.3cm}  &  22.860 \hspace{0.05cm}  \\ 
\hspace{0.05cm} 14.120 \hspace{0.05cm}  &  \hspace{0.05cm} -18.364 \hspace{0.3cm}  &  4.654  &  24.236  &  22.880 \hspace{0.3cm}  &  23.548 \hspace{0.05cm}  &  \hspace{0.05cm} -18.767 \hspace{0.3cm}  &  1.838  &  23.841  &  23.186 \hspace{0.3cm}  &  23.458 \hspace{0.05cm}  \\ 
\hline
\end{tabular}
} 
\end{center}
\end{table*}
\begin{table*}[!]
\begin{center}
\caption {Model D1M and $B_{12}=100.$} \label{tab:sigma_2}
\vskip 0.2cm
{\scriptsize
\begin{tabular}{ c | c c c c c | c c c c c }
\hline
&  &  &  $T$ = 1 MeV  &  &  &  &  &  $T$ = 5 MeV  &  &  \\ 
\hline
$\log_{10}\rho$ & $\log_{10}\tau$ & $\omega_c\tau$ & $\log_{10}\sigma$ & $\log_{10}\sigma_0$ & $\log_{10}\sigma_1$ & $\log_{10}\tau$ & $\omega_c\tau$ & $\log_{10}\sigma$ & $\log_{10}\sigma_0$ & $\log_{10}\sigma_1$ \\ 
\hline
\hspace{0.05cm} 11.825 \hspace{0.05cm}  &  \hspace{0.05cm} -19.421 \hspace{0.3cm}  &  1.285  &  22.186  &  21.759 \hspace{0.3cm}  &  21.870 \hspace{0.05cm}  &  \hspace{0.05cm} -19.477 \hspace{0.3cm}  &  1.128  &  22.182  &  21.757 \hspace{0.3cm}  &  21.861 \hspace{0.05cm}  \\ 
\hspace{0.05cm} 12.001 \hspace{0.05cm}  &  \hspace{0.05cm} -19.424 \hspace{0.3cm}  &  1.223  &  22.219  &  21.819 \hspace{0.3cm}  &  21.908 \hspace{0.05cm}  &  \hspace{0.05cm} -19.481 \hspace{0.3cm}  &  1.071  &  22.212  &  21.817 \hspace{0.3cm}  &  21.897 \hspace{0.05cm}  \\ 
\hspace{0.05cm} 12.167 \hspace{0.05cm}  &  \hspace{0.05cm} -19.427 \hspace{0.3cm}  &  1.162  &  22.255  &  21.882 \hspace{0.3cm}  &  21.949 \hspace{0.05cm}  &  \hspace{0.05cm} -19.485 \hspace{0.3cm}  &  1.016  &  22.246  &  21.879 \hspace{0.3cm}  &  21.935 \hspace{0.05cm}  \\ 
\hspace{0.05cm} 12.425 \hspace{0.05cm}  &  \hspace{0.05cm} -19.425 \hspace{0.3cm}  &  1.072  &  22.331  &  21.997 \hspace{0.3cm}  &  22.029 \hspace{0.05cm}  &  \hspace{0.05cm} -19.485 \hspace{0.3cm}  &  0.932  &  22.315  &  21.993 \hspace{0.3cm}  &  22.007 \hspace{0.05cm}  \\ 
\hspace{0.05cm} 12.795 \hspace{0.05cm}  &  \hspace{0.05cm} -19.400 \hspace{0.3cm}  &  0.965  &  22.495  &  22.208 \hspace{0.3cm}  &  22.194 \hspace{0.05cm}  &  \hspace{0.05cm} -19.468 \hspace{0.3cm}  &  0.826  &  22.464  &  22.201 \hspace{0.3cm}  &  22.156 \hspace{0.05cm}  \\ 
\hspace{0.05cm} 12.985 \hspace{0.05cm}  &  \hspace{0.05cm} -19.369 \hspace{0.3cm}  &  0.937  &  22.616  &  22.341 \hspace{0.3cm}  &  22.313 \hspace{0.05cm}  &  \hspace{0.05cm} -19.443 \hspace{0.3cm}  &  0.790  &  22.572  &  22.331 \hspace{0.3cm}  &  22.262 \hspace{0.05cm}  \\ 
\hspace{0.05cm} 13.173 \hspace{0.05cm}  &  \hspace{0.05cm} -19.307 \hspace{0.3cm}  &  0.963  &  22.775  &  22.489 \hspace{0.3cm}  &  22.474 \hspace{0.05cm}  &  \hspace{0.05cm} -19.393 \hspace{0.3cm}  &  0.790  &  22.715  &  22.479 \hspace{0.3cm}  &  22.404 \hspace{0.05cm}  \\ 
\hspace{0.05cm} 13.533 \hspace{0.05cm}  &  \hspace{0.05cm} - \hspace{0.3cm}  &  -  &  -  &  - \hspace{0.3cm}  &  - \hspace{0.05cm}  &  \hspace{0.05cm} -19.352 \hspace{0.3cm}  &  0.687  &  22.951  &  22.767 \hspace{0.3cm}  &  22.624 \hspace{0.05cm}  \\ 
\hspace{0.05cm} 13.700 \hspace{0.05cm}  &  \hspace{0.05cm} - \hspace{0.3cm}  &  -  &  -  &  - \hspace{0.3cm}  &  - \hspace{0.05cm}  &  \hspace{0.05cm} -19.198 \hspace{0.3cm}  &  0.871  &  23.205  &  22.944 \hspace{0.3cm}  &  22.899 \hspace{0.05cm}  \\ 
\hspace{0.05cm} 13.900 \hspace{0.05cm}  &  \hspace{0.05cm} - \hspace{0.3cm}  &  -  &  -  &  - \hspace{0.3cm}  &  - \hspace{0.05cm}  &  \hspace{0.05cm} -19.002 \hspace{0.3cm}  &  1.192  &  23.517  &  23.118 \hspace{0.3cm}  &  23.205 \hspace{0.05cm}  \\ 
\hspace{0.05cm} 14.001 \hspace{0.05cm}  &  \hspace{0.05cm} - \hspace{0.3cm}  &  -  &  -  &  - \hspace{0.3cm}  &  - \hspace{0.05cm}  &  \hspace{0.05cm} -18.907 \hspace{0.3cm}  &  1.383  &  23.669  &  23.191 \hspace{0.3cm}  &  23.340 \hspace{0.05cm}  \\ 
\hspace{0.05cm} 14.120 \hspace{0.05cm}  &  \hspace{0.05cm} - \hspace{0.3cm}  &  -  &  -  &  - \hspace{0.3cm}  &  - \hspace{0.05cm}  &  \hspace{0.05cm} -18.807 \hspace{0.3cm}  &  1.606  &  23.839  &  23.273 \hspace{0.3cm}  &  23.485 \hspace{0.05cm}  \\ 
\hline
\end{tabular}
} 
\end{center}
\end{table*}
\begin{table*}
\begin{center}
\caption {Model D1M* and $B_{12}=100.$} \label{tab:sigma_3}
\vskip 0.2cm
{\scriptsize
\begin{tabular}{ c | c c c c c | c c c c c }
\hline
&  &  &  $T$ = 1 MeV  &  &  &  &  &  $T$ = 5 MeV  &  &  \\ 
\hline
$\log_{10}\rho$ & $\log_{10}\tau$ & $\omega_c\tau$ & $\log_{10}\sigma$ & $\log_{10}\sigma_0$ & $\log_{10}\sigma_1$ & $\log_{10}\tau$ & $\omega_c\tau$ & $\log_{10}\sigma$ & $\log_{10}\sigma_0$ & $\log_{10}\sigma_1$ \\ 
\hline
\hspace{0.05cm} 11.825 \hspace{0.05cm}  &  \hspace{0.05cm} -19.402 \hspace{0.3cm}  &  1.349  &  22.199  &  21.745 \hspace{0.3cm}  &  21.878 \hspace{0.05cm}  &  \hspace{0.05cm} -19.461 \hspace{0.3cm}  &  1.178  &  22.196  &  21.744 \hspace{0.3cm}  &  21.868 \hspace{0.05cm}  \\ 
\hspace{0.05cm} 12.001 \hspace{0.05cm}  &  \hspace{0.05cm} -19.423 \hspace{0.3cm}  &  1.234  &  22.214  &  21.810 \hspace{0.3cm}  &  21.903 \hspace{0.05cm}  &  \hspace{0.05cm} -19.480 \hspace{0.3cm}  &  1.081  &  22.208  &  21.807 \hspace{0.3cm}  &  21.892 \hspace{0.05cm}  \\ 
\hspace{0.05cm} 12.167 \hspace{0.05cm}  &  \hspace{0.05cm} -19.426 \hspace{0.3cm}  &  1.172  &  22.250  &  21.872 \hspace{0.3cm}  &  21.943 \hspace{0.05cm}  &  \hspace{0.05cm} -19.484 \hspace{0.3cm}  &  1.025  &  22.241  &  21.870 \hspace{0.3cm}  &  21.929 \hspace{0.05cm}  \\ 
\hspace{0.05cm} 12.425 \hspace{0.05cm}  &  \hspace{0.05cm} -19.424 \hspace{0.3cm}  &  1.081  &  22.325  &  21.986 \hspace{0.3cm}  &  22.022 \hspace{0.05cm}  &  \hspace{0.05cm} -19.485 \hspace{0.3cm}  &  0.941  &  22.309  &  21.983 \hspace{0.3cm}  &  22.002 \hspace{0.05cm}  \\ 
\hspace{0.05cm} 12.795 \hspace{0.05cm}  &  \hspace{0.05cm} -19.401 \hspace{0.3cm}  &  0.972  &  22.487  &  22.196 \hspace{0.3cm}  &  22.185 \hspace{0.05cm}  &  \hspace{0.05cm} -19.468 \hspace{0.3cm}  &  0.833  &  22.456  &  22.189 \hspace{0.3cm}  &  22.148 \hspace{0.05cm}  \\ 
\hspace{0.05cm} 12.985 \hspace{0.05cm}  &  \hspace{0.05cm} -19.366 \hspace{0.3cm}  &  0.953  &  22.609  &  22.327 \hspace{0.3cm}  &  22.307 \hspace{0.05cm}  &  \hspace{0.05cm} -19.440 \hspace{0.3cm}  &  0.804  &  22.566  &  22.318 \hspace{0.3cm}  &  22.256 \hspace{0.05cm}  \\ 
\hspace{0.05cm} 13.173 \hspace{0.05cm}  &  \hspace{0.05cm} -19.307 \hspace{0.3cm}  &  0.977  &  22.764  &  22.472 \hspace{0.3cm}  &  22.463 \hspace{0.05cm}  &  \hspace{0.05cm} -19.392 \hspace{0.3cm}  &  0.802  &  22.705  &  22.462 \hspace{0.3cm}  &  22.395 \hspace{0.05cm}  \\ 
\hspace{0.05cm} 13.533 \hspace{0.05cm}  &  \hspace{0.05cm} -19.069 \hspace{0.3cm}  &  1.340  &  23.203  &  22.755 \hspace{0.3cm}  &  22.883 \hspace{0.05cm}  &  \hspace{0.05cm} -19.206 \hspace{0.3cm}  &  0.978  &  23.083  &  22.771 \hspace{0.3cm}  &  22.779 \hspace{0.05cm}  \\ 
\hspace{0.05cm} 13.700 \hspace{0.05cm}  &  \hspace{0.05cm} -18.891 \hspace{0.3cm}  &  1.810  &  23.476  &  22.845 \hspace{0.3cm}  &  23.103 \hspace{0.05cm}  &  \hspace{0.05cm} -19.077 \hspace{0.3cm}  &  1.178  &  23.304  &  22.907 \hspace{0.3cm}  &  22.992 \hspace{0.05cm}  \\ 
\hspace{0.05cm} 13.900 \hspace{0.05cm}  &  \hspace{0.05cm} -18.666 \hspace{0.3cm}  &  2.668  &  23.814  &  22.903 \hspace{0.3cm}  &  23.330 \hspace{0.05cm}  &  \hspace{0.05cm} -18.930 \hspace{0.3cm}  &  1.451  &  23.561  &  23.051 \hspace{0.3cm}  &  23.223 \hspace{0.05cm}  \\ 
\hspace{0.05cm} 14.001 \hspace{0.05cm}  &  \hspace{0.05cm} -18.541 \hspace{0.3cm}  &  3.324  &  23.996  &  22.915 \hspace{0.3cm}  &  23.437 \hspace{0.05cm}  &  \hspace{0.05cm} -18.855 \hspace{0.3cm}  &  1.613  &  23.692  &  23.121 \hspace{0.3cm}  &  23.337 \hspace{0.05cm}  \\ 
\hspace{0.05cm} 14.068 \hspace{0.05cm}  &  \hspace{0.05cm} -18.460 \hspace{0.3cm}  &  3.828  &  24.119  &  22.924 \hspace{0.3cm}  &  23.507 \hspace{0.05cm}  &  \hspace{0.05cm} -18.809 \hspace{0.3cm}  &  1.714  &  23.779  &  23.170 \hspace{0.3cm}  &  23.411 \hspace{0.05cm}  \\ 
\hspace{0.05cm} 14.120 \hspace{0.05cm}  &  \hspace{0.05cm} - \hspace{0.3cm}  &  -  &  -  &  - \hspace{0.3cm}  &  - \hspace{0.05cm}  &  \hspace{0.05cm} -18.735 \hspace{0.3cm}  &  1.947  &  23.889  &  23.195 \hspace{0.3cm}  &  23.491 \hspace{0.05cm}  \\ 
\hline
\end{tabular}
} 
\end{center}
\end{table*}
\begin{table*}
\begin{center}
\caption {Model Bsk24 and $B_{12}=100.$} \label{tab:sigma_4}
\vskip 0.2cm
{\scriptsize
\begin{tabular}{ c | c c c c c | c c c c c }
\hline
&  &  &  $T$ = 1 MeV  &  &  &  &  &  $T$ = 5 MeV  &  &  \\ 
\hline
$\log_{10}\rho$ & $\log_{10}\tau$ & $\omega_c\tau$ & $\log_{10}\sigma$ & $\log_{10}\sigma_0$ & $\log_{10}\sigma_1$ & $\log_{10}\tau$ & $\omega_c\tau$ & $\log_{10}\sigma$ & $\log_{10}\sigma_0$ & $\log_{10}\sigma_1$ \\ 
\hline
\hspace{0.05cm} 11.655 \hspace{0.05cm}  &  \hspace{0.05cm} -19.514 \hspace{0.3cm}  &  1.054  &  22.078  &  21.752 \hspace{0.3cm}  &  21.776 \hspace{0.05cm}  &  \hspace{0.05cm} -19.560 \hspace{0.3cm}  &  0.947  &  22.067  &  21.750 \hspace{0.3cm}  &  21.762 \hspace{0.05cm}  \\ 
\hspace{0.05cm} 11.962 \hspace{0.05cm}  &  \hspace{0.05cm} -19.538 \hspace{0.3cm}  &  0.923  &  22.119  &  21.850 \hspace{0.3cm}  &  21.817 \hspace{0.05cm}  &  \hspace{0.05cm} -19.584 \hspace{0.3cm}  &  0.831  &  22.106  &  21.847 \hspace{0.3cm}  &  21.800 \hspace{0.05cm}  \\ 
\hspace{0.05cm} 12.320 \hspace{0.05cm}  &  \hspace{0.05cm} -19.568 \hspace{0.3cm}  &  0.779  &  22.176  &  21.969 \hspace{0.3cm}  &  21.862 \hspace{0.05cm}  &  \hspace{0.05cm} -19.613 \hspace{0.3cm}  &  0.702  &  22.160  &  21.963 \hspace{0.3cm}  &  21.840 \hspace{0.05cm}  \\ 
\hspace{0.05cm} 12.627 \hspace{0.05cm}  &  \hspace{0.05cm} -19.598 \hspace{0.3cm}  &  0.649  &  22.245  &  22.092 \hspace{0.3cm}  &  21.905 \hspace{0.05cm}  &  \hspace{0.05cm} -19.643 \hspace{0.3cm}  &  0.586  &  22.226  &  22.082 \hspace{0.3cm}  &  21.877 \hspace{0.05cm}  \\ 
\hspace{0.05cm} 12.934 \hspace{0.05cm}  &  \hspace{0.05cm} -19.628 \hspace{0.3cm}  &  0.525  &  22.340  &  22.234 \hspace{0.3cm}  &  21.955 \hspace{0.05cm}  &  \hspace{0.05cm} -19.672 \hspace{0.3cm}  &  0.474  &  22.318  &  22.220 \hspace{0.3cm}  &  21.921 \hspace{0.05cm}  \\ 
\hspace{0.05cm} 13.240 \hspace{0.05cm}  &  \hspace{0.05cm} -19.644 \hspace{0.3cm}  &  0.426  &  22.472  &  22.399 \hspace{0.3cm}  &  22.030 \hspace{0.05cm}  &  \hspace{0.05cm} -19.689 \hspace{0.3cm}  &  0.384  &  22.446  &  22.380 \hspace{0.3cm}  &  21.986 \hspace{0.05cm}  \\ 
\hspace{0.05cm} 13.599 \hspace{0.05cm}  &  \hspace{0.05cm} -19.610 \hspace{0.3cm}  &  0.371  &  22.695  &  22.639 \hspace{0.3cm}  &  22.209 \hspace{0.05cm}  &  \hspace{0.05cm} -19.660 \hspace{0.3cm}  &  0.331  &  22.659  &  22.610 \hspace{0.3cm}  &  22.148 \hspace{0.05cm}  \\ 
\hspace{0.05cm} 13.905 \hspace{0.05cm}  &  \hspace{0.05cm} -19.459 \hspace{0.3cm}  &  0.436  &  23.008  &  22.933 \hspace{0.3cm}  &  22.572 \hspace{0.05cm}  &  \hspace{0.05cm} -19.529 \hspace{0.3cm}  &  0.371  &  22.949  &  22.890 \hspace{0.3cm}  &  22.472 \hspace{0.05cm}  \\ 
\hspace{0.05cm} 14.047 \hspace{0.05cm}  &  \hspace{0.05cm} -19.244 \hspace{0.3cm}  &  0.650  &  23.307  &  23.154 \hspace{0.3cm}  &  22.966 \hspace{0.05cm}  &  \hspace{0.05cm} -19.351 \hspace{0.3cm}  &  0.507  &  23.209  &  23.104 \hspace{0.3cm}  &  22.821 \hspace{0.05cm}  \\ 
\hspace{0.05cm} 14.056 \hspace{0.05cm}  &  \hspace{0.05cm} -19.220 \hspace{0.3cm}  &  0.680  &  23.337  &  23.172 \hspace{0.3cm}  &  23.004 \hspace{0.05cm}  &  \hspace{0.05cm} -19.333 \hspace{0.3cm}  &  0.525  &  23.235  &  23.123 \hspace{0.3cm}  &  22.855 \hspace{0.05cm}  \\ 
\hspace{0.05cm} 14.063 \hspace{0.05cm}  &  \hspace{0.05cm} -19.199 \hspace{0.3cm}  &  0.711  &  23.362  &  23.185 \hspace{0.3cm}  &  23.037 \hspace{0.05cm}  &  \hspace{0.05cm} -19.316 \hspace{0.3cm}  &  0.543  &  23.255  &  23.137 \hspace{0.3cm}  &  22.883 \hspace{0.05cm}  \\ 
\hspace{0.05cm} 14.085 \hspace{0.05cm}  &  \hspace{0.05cm} -19.024 \hspace{0.3cm}  &  1.181  &  23.448  &  23.069 \hspace{0.3cm}  &  23.141 \hspace{0.05cm}  &  \hspace{0.05cm} -19.177 \hspace{0.3cm}  &  0.829  &  23.306  &  23.067 \hspace{0.3cm}  &  22.998 \hspace{0.05cm}  \\ 
\hspace{0.05cm} 14.098 \hspace{0.05cm}  &  \hspace{0.05cm} -18.934 \hspace{0.3cm}  &  1.490  &  23.517  &  23.008 \hspace{0.3cm}  &  23.182 \hspace{0.05cm}  &  \hspace{0.05cm} -19.111 \hspace{0.3cm}  &  0.991  &  23.351  &  23.040 \hspace{0.3cm}  &  23.048 \hspace{0.05cm}  \\ 
\hline
\end{tabular}
} 
\end{center}
\end{table*}
\begin{table*}[!]
\begin{center}
\caption {Model Sly9 and $B_{12}=100$.} \label{tab:sigma_5}
\vskip 0.2cm
{\scriptsize
\begin{tabular}{ c | c c c c c | c c c c c }
\hline
&  &  &  $T$ = 1 MeV  &  &  &  &  &  $T$ = 5 MeV  &  &  \\ 
\hline
$\log_{10}\rho$ & $\log_{10}\tau$ & $\omega_c\tau$ & $\log_{10}\sigma$ & $\log_{10}\sigma_0$ & $\log_{10}\sigma_1$ & $\log_{10}\tau$ & $\omega_c\tau$ & $\log_{10}\sigma$ & $\log_{10}\sigma_0$ & $\log_{10}\sigma_1$ \\ 
\hline
\hspace{0.05cm} 11.613 \hspace{0.05cm}  &  \hspace{0.05cm} -19.464 \hspace{0.3cm}  &  1.190  &  22.121  &  21.736 \hspace{0.3cm}  &  21.813 \hspace{0.05cm}  &  \hspace{0.05cm} -19.516 \hspace{0.3cm}  &  1.057  &  22.106  &  21.736 \hspace{0.3cm}  &  21.798 \hspace{0.05cm}  \\ 
\hspace{0.05cm} 12.123 \hspace{0.05cm}  &  \hspace{0.05cm} -19.510 \hspace{0.3cm}  &  0.954  &  22.176  &  21.894 \hspace{0.3cm}  &  21.875 \hspace{0.05cm}  &  \hspace{0.05cm} -19.559 \hspace{0.3cm}  &  0.852  &  22.160  &  21.891 \hspace{0.3cm}  &  21.855 \hspace{0.05cm}  \\ 
\hspace{0.05cm} 12.623 \hspace{0.05cm}  &  \hspace{0.05cm} -19.555 \hspace{0.3cm}  &  0.738  &  22.263  &  22.073 \hspace{0.3cm}  &  21.942 \hspace{0.05cm}  &  \hspace{0.05cm} -19.603 \hspace{0.3cm}  &  0.661  &  22.243  &  22.064 \hspace{0.3cm}  &  21.915 \hspace{0.05cm}  \\ 
\hspace{0.05cm} 12.891 \hspace{0.05cm}  &  \hspace{0.05cm} -19.576 \hspace{0.3cm}  &  0.632  &  22.334  &  22.188 \hspace{0.3cm}  &  21.990 \hspace{0.05cm}  &  \hspace{0.05cm} -19.624 \hspace{0.3cm}  &  0.566  &  22.312  &  22.176 \hspace{0.3cm}  &  21.956 \hspace{0.05cm}  \\ 
\hspace{0.05cm} 13.091 \hspace{0.05cm}  &  \hspace{0.05cm} -19.582 \hspace{0.3cm}  &  0.569  &  22.408  &  22.285 \hspace{0.3cm}  &  22.041 \hspace{0.05cm}  &  \hspace{0.05cm} -19.631 \hspace{0.3cm}  &  0.508  &  22.382  &  22.270 \hspace{0.3cm}  &  22.002 \hspace{0.05cm}  \\ 
\hspace{0.05cm} 13.191 \hspace{0.05cm}  &  \hspace{0.05cm} -19.580 \hspace{0.3cm}  &  0.545  &  22.452  &  22.339 \hspace{0.3cm}  &  22.076 \hspace{0.05cm}  &  \hspace{0.05cm} -19.630 \hspace{0.3cm}  &  0.486  &  22.424  &  22.321 \hspace{0.3cm}  &  22.032 \hspace{0.05cm}  \\ 
\hspace{0.05cm} 13.291 \hspace{0.05cm}  &  \hspace{0.05cm} -19.571 \hspace{0.3cm}  &  0.529  &  22.504  &  22.397 \hspace{0.3cm}  &  22.121 \hspace{0.05cm}  &  \hspace{0.05cm} -19.623 \hspace{0.3cm}  &  0.469  &  22.473  &  22.377 \hspace{0.3cm}  &  22.072 \hspace{0.05cm}  \\ 
\hspace{0.05cm} 13.393 \hspace{0.05cm}  &  \hspace{0.05cm} -19.555 \hspace{0.3cm}  &  0.520  &  22.565  &  22.461 \hspace{0.3cm}  &  22.178 \hspace{0.05cm}  &  \hspace{0.05cm} -19.610 \hspace{0.3cm}  &  0.459  &  22.530  &  22.438 \hspace{0.3cm}  &  22.123 \hspace{0.05cm}  \\ 
\hspace{0.05cm} 13.491 \hspace{0.05cm}  &  \hspace{0.05cm} -19.530 \hspace{0.3cm}  &  0.524  &  22.634  &  22.529 \hspace{0.3cm}  &  22.248 \hspace{0.05cm}  &  \hspace{0.05cm} -19.589 \hspace{0.3cm}  &  0.457  &  22.594  &  22.503 \hspace{0.3cm}  &  22.185 \hspace{0.05cm}  \\ 
\hspace{0.05cm} 13.591 \hspace{0.05cm}  &  \hspace{0.05cm} -19.493 \hspace{0.3cm}  &  0.542  &  22.717  &  22.605 \hspace{0.3cm}  &  22.340 \hspace{0.05cm}  &  \hspace{0.05cm} -19.558 \hspace{0.3cm}  &  0.467  &  22.670  &  22.576 \hspace{0.3cm}  &  22.265 \hspace{0.05cm}  \\ 
\hspace{0.05cm} 13.691 \hspace{0.05cm}  &  \hspace{0.05cm} -19.442 \hspace{0.3cm}  &  0.578  &  22.813  &  22.688 \hspace{0.3cm}  &  22.450 \hspace{0.05cm}  &  \hspace{0.05cm} -19.516 \hspace{0.3cm}  &  0.488  &  22.756  &  22.655 \hspace{0.3cm}  &  22.362 \hspace{0.05cm}  \\ 
\hspace{0.05cm} 13.791 \hspace{0.05cm}  &  \hspace{0.05cm} -19.377 \hspace{0.3cm}  &  0.639  &  22.922  &  22.774 \hspace{0.3cm}  &  22.579 \hspace{0.05cm}  &  \hspace{0.05cm} -19.462 \hspace{0.3cm}  &  0.525  &  22.852  &  22.738 \hspace{0.3cm}  &  22.475 \hspace{0.05cm}  \\ 
\hspace{0.05cm} 13.809 \hspace{0.05cm}  &  \hspace{0.05cm} -19.364 \hspace{0.3cm}  &  0.653  &  22.943  &  22.789 \hspace{0.3cm}  &  22.604 \hspace{0.05cm}  &  \hspace{0.05cm} -19.451 \hspace{0.3cm}  &  0.534  &  22.871  &  22.753 \hspace{0.3cm}  &  22.497 \hspace{0.05cm}  \\ 
\hline
\end{tabular}
} 
\end{center}
\end{table*}
\end{widetext}
\end{document}